\documentclass[12pt]{article}
\usepackage{latexsym,amssymb,amsfonts,amsmath}
\usepackage[dvips]{graphicx}
\newlength{\extraspace}
\newlength{\extraspaces}
\newcommand{\be}{\begin{equation}
\addtolength{\abovedisplayskip}{\extraspaces}
\addtolength{\belowdisplayskip}{\extraspaces}
\addtolength{\abovedisplayshortskip}{\extraspace}
\addtolength{\belowdisplayshortskip}{\extraspace}}
\newcommand{\ee}{\end{equation}}

\newcommand{\ba}{\begin{eqnarray}
\addtolength{\abovedisplayskip}{\extraspaces}
\addtolength{\belowdisplayskip}{\extraspaces}
\addtolength{\abovedisplayshortskip}{\extraspace}
\addtolength{\belowdisplayshortskip}{\extraspace}}
\newcommand{\ea}{\end{eqnarray}}

\newcommand{\newsection}[1]{
\vspace{15mm}
\pagebreak[3]
\addtocounter{section}{1}
\setcounter{equation}{0}
\setcounter{subsection}{0}
\setcounter{footnote}{0}
\begin{flushleft}
{\Large\bf \thesection. #1}
\end{flushleft}
\nopagebreak
\medskip
\nopagebreak}

\def\Tr{{\rm Tr}}
\def\re{{\rm Re}}
\def\im{{\rm Im}}

\begin{document}

\addtolength{\baselineskip}{.8mm}

{\thispagestyle{empty}

\noindent \hspace{1cm}  \hfill DFUP--TH/2006--5 \hspace{1cm}\\
\mbox{}                 \hfill April 2006 \hspace{1cm}\\

\begin{center}
\vspace*{1.0cm}
{\large\bf ANALYTICITY AND CROSSING SYMMETRY OF THE} \\
{\large\bf EIKONAL AMPLITUDES IN GAUGE THEORIES} \\
\vspace*{1.0cm}
{\large Matteo Giordano, ~Enrico Meggiolaro\footnote{E--mail:
enrico.meggiolaro@df.unipi.it} }\\
\vspace*{0.5cm}{\normalsize
{Dipartimento di Fisica,
Universit\`a di Pisa,\\
Largo Pontecorvo 3,
I--56127 Pisa, Italy.}}\\
\vspace*{2cm}{\large \bf Abstract}
\end{center}

\noindent
After a brief review and a more refined analysis of some relevant analyticity
properties (when going from Minkowskian to Euclidean theory) of the
high--energy parton--parton and hadron--hadron scattering amplitudes in gauge
theories, described nonperturbatively, in the eikonal approximation, by
certain correlation functions of two Wilson lines or two Wilson loops near
the light cone, we shall see how these same properties lead to a nice
geometrical interpretation of the crossing symmetry between quark--quark and
quark--antiquark eikonal amplitudes and also between loop--loop eikonal
amplitudes. This relation between Minkowskian--to--Euclidean
analyticity properties and crossing symmetry is discussed in detail and
explicitly tested in the first orders of perturbation theory.
Some nonperturbative examples existing in the literature are also discussed.
\\
}
\newpage

\newsection{Introduction}

\noindent
A big effort has been made in the last fifteen years (since the seminal
paper by Nachtmann in 1991 \cite{Nachtmann91}) in the nonperturbative study,
from the first principles of QCD, of the high--energy parton--parton and
hadron--hadron elastic scattering amplitudes (for a review, see Refs.
\cite{Dosch,pomeron-book}): these can be described, in the so--called
{\it eikonal} approximation (and, therefore, they will be sometimes called
{\it ``eikonal scattering amplitudes''}), by certain correlation functions
of two Wilson lines or two Wilson loops near the light cone.

The section 2 of this paper contains a brief review (for the benefit of the
reader) and a more refined analysis of some relevant analyticity properties
of the line--line and loop--loop correlation functions in gauge theories,
when going from Minkowskian to Euclidean theory: these properties make it
possible to reconstruct the eikonal scattering amplitudes starting from the
Euclidean correlation functions, which can be computed with nonperturbative
techniques (some examples existing in the literature will be discussed in
section 5).

In section 3 we will show (always in a nonperturbative way, using the
functional integral approach) how these same properties also lead to a nice
geometrical interpretation of the {\it crossing symmetry} between quark--quark
and quark--antiquark correlators and also between loop--loop correlation
functions. This relation between Minkowskian--to--Euclidean analyticity
properties and crossing symmetry is the main novel result of this paper
and is discussed in detail also in the two last sections of it.

In particular, in section 4 (and appendix A) it is explicitly tested in the
first orders of perturbation theory, which is the only available technique for
computing (from first principles) {\it both} the Minkowskian {\it and} the
Euclidean line--line and loop--loop correlation functions.
As already stressed in Ref. \cite{Meggiolaro97} (but see also Refs.
\cite{Balitsky,BB} and references therein), such perturbative expansions of the
line--line and loop--loop correlation functions, when considered in the
Minkowskian theory in the limit of very large rapidity gap, must be
eventually compared (as a non--trivial check!) to the well--known results
obtained when computing the high--energy scattering amplitudes with usual
perturbative techniques \cite{BFKL,Lipatov,Cheng-Wu-book}.

Finally, some nonperturbative examples existing in the literature and also the
necessity of a real nonperturbative foundation of the above--mentioned
analyticity properties are discussed as concluding remarks in section 5
(and appendix B), together with some prospects for the future.

\newsection{Eikonal scattering amplitudes}

\noindent
The parton--parton elastic scattering amplitude, at high squared energies $s$
in the center of mass and small squared transferred momentum $t$
(that is to say: $|t| \le 1~{\rm GeV}^2 \ll s$),
can be described by the expectation value of two {\it infinite lightlike}
Wilson lines, running along the classical trajectories of the colliding
particles \cite{Nachtmann91,Verlinde,Korchemsky,Meggiolaro96,Meggiolaro01}.
However, this description is affected by infrared (IR) divergences
\cite{Verlinde,Korchemsky}, which are typical of $3 + 1$ dimensional gauge
theories. One can regularize this IR problem by letting the Wilson lines
coincide with the classical trajectories for partons with a non--zero mass $m$
(so forming a certain {\it finite} rapidity gap, i.e., a certain {\it finite}
hyperbolic angle $\chi$ in Minkowskian space--time: of course [see Eq.
(\ref{s-chi}) below], $\chi \simeq \log(s/m^2) \to \infty$ when $s \to \infty$)
and, in addition, by considering {\it finite} Wilson lines, extending in
proper time from $-T$ to $T$ (and eventually letting $T \to +\infty$)
\cite{Meggiolaro97,Verlinde,Korchemsky,Meggiolaro98,Meggiolaro02}.
For example, the high--energy quark--quark
elastic scattering amplitude ${\cal M}^{qq}(s,t)$ is (explicitly indicating
the colour indices $i,j$ [initial] and $i',j'$ [final] and the spin indices
$\alpha,\beta$ [initial] and $\alpha',\beta'$ [final] of the colliding quarks):
\be
\mathcal{M}^{qq}(s;t)_{i'i;j'j}^{\alpha'\alpha;\beta'\beta}
\mathop{\sim}_{s \to \infty}
-i~ 2s~ \delta_{\alpha'\alpha} \delta_{\beta'\beta}~
g_M^{qq}(p_1,p_2;T\to\infty;t)_{i'i;j'j} ,
\label{scatt}
\ee
with $g_M^{qq}$ defined as:
\be
g_M^{qq} (p_1,p_2;T;t)_{i'i;j'j} \equiv {1 \over [Z_M(T)]^2}
\displaystyle\int d^2 \vec{z}_\perp e^{i \vec{q}_\perp \cdot \vec{z}_\perp}
\langle [ W^{(T)}_{p_1} (\vec{z}_\perp) - \mathbb{I} ]_{i'i}
[ W^{(T)}_{p_2} (\vec{0}_\perp) - \mathbb{I} ]_{j'j} \rangle ,
\label{gM}
\ee
where $t = -|\vec{q}_\perp|^2$, $\vec{q}_\perp$ being the tranferred momentum,
and $\vec{z}_\perp = (z^2,z^3)$ is the distance between the two trajectories
in the {\it transverse} plane ({\it impact parameter}). We are taking the two
colliding quarks (with mass $m$) moving (in the center--of--mass system) with
speed $V$ and $-V$ along, for example, the $x^1$--direction and so having
four--momenta $p_1$ and $p_2$ given by:
\ba
p_1 &=& m (\cosh {\chi \over 2},\sinh {\chi \over 2},\vec{0}_\perp) ,
\nonumber \\
p_2 &=& m (\cosh {\chi \over 2},-\sinh {\chi \over 2},\vec{0}_\perp) ,
\label{p1p2}
\ea
where $\chi = 2~{\rm arctanh} V$ is the hyperbolic angle between the two
trajectories (i.e., $p_1 \cdot p_2 = m^2 \cosh\chi$). Therefore:
\be
s \equiv (p_1 + p_2)^2 = 2 m^2 ( \cosh \chi + 1 ) .
\label{s-chi}
\ee
The two IR--regularized Wilson lines are defined as
[$z = (0,0,\vec{z}_\perp)$]:
\ba
W^{(T)}_{p_1} (\vec{z}_\perp) &\equiv&
{\cal T} \exp \left[ -ig \displaystyle\int_{-T}^{+T}
A_\mu (z + {p_1 \over m} \tau) {p_1^\mu \over m} d\tau \right] ,
\nonumber \\
W^{(T)}_{p_2} (\vec{0}_\perp) &\equiv&
{\cal T} \exp \left[ -ig \displaystyle\int_{-T}^{+T}
A_\mu ({p_2 \over m} \tau) {p_2^\mu \over m} d\tau \right] ,
\label{linesM}
\ea
where ${\cal T}$ stands for ``{\it time ordering}'' and, for a non--Abelian
gauge theory with $N_c$ colours, $A_\mu = A_\mu^a T_a$, $T^a$ ($a = 1, \ldots,
N_c^2 - 1$) being the generators of the $SU(N_c)$ algebra in the fundamental
representation.
The two Wilson lines are schematically shown in Fig. 1.\\
Finally, $Z_M(T)$ is a sort of Wilson--line renormalization constant:
\be
Z_M(T) \equiv {1 \over N_c} \langle \Tr [ W^{(T)}_{p_1} (\vec{0}_\perp) ]
\rangle = {1 \over N_c} \langle \Tr [ W^{(T)}_{p_2} (\vec{0}_\perp) ] \rangle .
\label{ZW}
\ee
The expectation values $\langle W_{p_1} W_{p_2} \rangle$,
$\langle W_{p_1} \rangle$ and $\langle W_{p_2} \rangle$ are averages in the
sense of the QCD functional integrals:
\ba
\langle {\cal O}[A] \rangle &=&
{1 \over Z} \displaystyle\int [dA] \det(Q[A]) e^{iS_A} {\cal O}[A] ,
\nonumber \\
Z &=& \displaystyle\int [dA] \det(Q[A]) e^{iS_A} ,
\label{fintM}
\ea
where $S_A$ is the pure--gauge (Yang--Mills) action and $Q[A]$ is the
{\it quark matrix}, coming from the functional integration over the fermion
degrees of freedom.

The correlation function (\ref{gM}), with the four--vectors $p_1$ and
$p_2$ defined by Eq. (\ref{p1p2}), will be also denoted (with a slight
abuse of notation) as:\footnote{We remark that only the {\it asymptotic}
behaviour for $\chi \simeq \log(s/m^2) \to \infty$ of the correlator $g_M^{qq}$
describes the high--energy quark--quark elastic scattering amplitude
by virtue of Eq. (\ref{scatt}). The correlator $g_M^{qq}$ as a function
of the generic hyperbolic angle $\chi$ between the two Wilson lines,
defined by Eqs. (\ref{gM}) and (\ref{p1p2}), must not
be identified with the scattering amplitude at {\it every} $\chi$, i.e.,
at {\it every} $s = 2 m^2 ( \cosh \chi + 1 )$.}
\be
g_M^{qq} (p_1,p_2;T;t)_{i'i;j'j} \equiv g_M^{qq} (\chi;T;t)_{i'i;j'j} .
\ee
By virtue of the invariance under parity transformations and $O(3)$ spatial
rotations, the domain of the function $g_M$ in the variable $\chi$ can be
restricted to the real positive axis, $\chi \in \mathbb{R}^+$.
In fact, a parity transformation together with a $180^\circ$ rotation
around the $x^1$ axis, i.e., a transformation
\be
\label{eq:trasfoM}
x \to x' = \Lambda x,~~ \qquad \Lambda = \left( 
\begin{array}{cccc}
1 & 0 & 0 & 0\\
0 & -1 & 0 & 0\\
0 & 0 & 1 & 0\\
0 & 0 & 0 & 1
\end{array}\right) ,
\ee
brings $\chi$ into $-\chi$ without modifying the
functional integral:
\be
g_M^{qq} (-\chi;T;t)_{i'i;j'j} = g_M^{qq} (\chi;T;t)_{i'i;j'j} ,
\qquad \forall \chi\in\mathbb{R} .
\label{propM}
\ee
Turning now to the Euclidean theory, one can consider
the corresponding quantity $g_E^{qq}$, defined as a
(properly normalized) correlation function of two (IR--regularized) Euclidean
Wilson lines $\widetilde{W}_{p_{1E}}$ and $\widetilde{W}_{p_{2E}}$, i.e.,
\ba
g_E^{qq} (p_{1E},p_{2E};T;t)_{i'i;j'j} &\equiv& {1 \over [Z_E(T)]^2}
\displaystyle\int d^2 \vec{z}_\perp e^{i \vec{q}_\perp \cdot \vec{z}_\perp}
\langle [ \widetilde{W}^{(T)}_{p_{1E}} (\vec{z}_\perp) - \mathbb{I} ]_{i'i}
[ \widetilde{W}^{(T)}_{p_{2E}} (\vec{0}_\perp) - \mathbb{I} ]_{j'j} \rangle_E ,
\nonumber \\
Z_E(T) &\equiv& {1 \over N_c}
\langle \Tr [ \widetilde{W}^{(T)}_{p_{1E}} (\vec{0}_\perp) ] \rangle_E =
{1 \over N_c} \langle \Tr [ \widetilde{W}^{(T)}_{p_{2E}} (\vec{0}_\perp) ]
\rangle_E ,
\label{gE}
\ea
where [$z_E = (0,\vec{z}_\perp,0)$]:
\ba
\widetilde{W}^{(T)}_{p_{1E}} (\vec{z}_\perp) &\equiv&
{\cal T} \exp \left[ -ig \displaystyle\int_{-T}^{+T}
A^{(E)}_\mu (z_E + {p_{1E} \over m} \tau) {p_{1E\mu} \over m} d\tau \right] ,
\nonumber \\
\widetilde{W}^{(T)}_{p_{2E}} (\vec{0}_\perp) &\equiv&
{\cal T} \exp \left[ -ig \displaystyle\int_{-T}^{+T}
A^{(E)}_\mu ({p_{2E} \over m} \tau) {p_{2E\mu} \over m} d\tau \right] ,
\label{linesE}
\ea
and:
\ba
\langle {\cal O}[A^{(E)}] \rangle_E &=&
{1 \over Z^{(E)}} \displaystyle\int [dA^{(E)}] \det(Q^{(E)}[A^{(E)}])
e^{-S^{(E)}_A} {\cal O}[A^{(E)}] ,\nonumber \\
Z^{(E)} &=& \displaystyle\int [dA^{(E)}]
\det(Q^{(E)}[A^{(E)}]) e^{-S^{(E)}_A} ,
\label{fintE}
\ea
$S^{(E)}_A$ being the Euclidean pure--gauge (Yang--Mills) action and
$Q^{(E)}[A]$ being the Euclidean {\it quark matrix}, coming from the
functional integration over the fermion degrees of freedom.
The two Euclidean four--vectors $p_{1E}$ and $p_{2E}$ are chosen to be:
\ba
p_{1E} &=& m (\sin{\theta \over 2}, \vec{0}_\perp, \cos{\theta \over 2} ) ,
\nonumber \\
p_{2E} &=& m (-\sin{\theta \over 2}, \vec{0}_\perp, \cos{\theta \over 2} ) ,
\label{p1p2E}
\ea
$\theta$ being the angle formed by the two trajectories in the Euclidean
four--space (i.e., $p_{1E} \cdot p_{2E} = m^2 \cos\theta$).

The correlation function $g_E$ in (\ref{gE}), with the four--vectors $p_{1E}$
and $p_{2E}$ defined by Eq. (\ref{p1p2E}), will be also denoted (with a slight
abuse of notation) as:
\be
g_E^{qq} (p_{1E},p_{2E};T;t)_{i'i;j'j} \equiv
g_E^{qq} (\theta;T;t)_{i'i;j'j} .
\ee
By virtue of the $O(4)$ symmetry of the Euclidean theory,
the domain of the function $g_E$ in the variable $\theta$ can be
restricted to the interval $(0,\pi)$.
In fact, the invariance of the functional integral under the following
$O(4)$ transformation:
\be
\label{eq:trasfoE1}
x_E \to x'_E = {\cal R}_1 x_E,~~ \qquad {\cal R}_1 = \left( 
\begin{array}{cccc}
-1 & 0 & 0 & 0\\
0 & 1 & 0 & 0\\
0 & 0 & 1 & 0\\
0 & 0 & 0 & 1
\end{array}\right) ,
\ee
leads to the following relation:
\be
g_E^{qq} (-\theta;T;t)_{i'i;j'j} = g_E^{qq} (\theta;T;t)_{i'i;j'j} ,
\qquad \forall\theta\in\mathbb{R}.
\label{propE1}
\ee
Similarly, the invariance of the functional integral under the following
$O(4)$ transformation:
\be
\label{eq:trasfoE2}
x_E \to x'_E = {\cal R}_2 x_E,~~ \qquad {\cal R}_2 = \left( 
\begin{array}{cccc}
1 & 0 & 0 & 0\\
0 & 1 & 0 & 0\\
0 & 0 & 1 & 0\\
0 & 0 & 0 & -1
\end{array}\right) ,
\ee
leads to the following relation:
\be
g_E^{qq} (2\pi-\theta;T;t)_{i'i;j'j} = g_E^{qq} (\theta;T;t)_{i'i;j'j} ,
\qquad \forall\theta\in\mathbb{R}.
\label{propE2}
\ee
These two relations imply the possibility of restricting the domain in the
angular variable $\theta$ to the interval $(0,\pi)$,
as said above.\footnote{After substituting $\theta \to \theta + 2\pi$ into
Eq. (\ref{propE2}) and using also Eq. (\ref{propE1}), one finds that
$g_E(\theta+2\pi;T;t) = g_E(-\theta;T;t) = g_E(\theta;T;t)$.
Moreover, after substituting $\theta \to 2\pi - \theta$ into Eq. (\ref{propE1})
and using also Eq. (\ref{propE2}), one finds that
$g_E(\theta-2\pi;T;t) = g_E(2\pi-\theta;T;t) = g_E(\theta;T;t)$.
Therefore we conclude that:
$g_E(\theta+2\pi k;T;t) = g_E(\theta;T;t)$, $\forall k\in\mathbb{Z}$.}

The quantity $g_M^{qq} (\chi;T;t)$ with $\chi\in\mathbb{R}^+$ can be
reconstructed from the corresponding Euclidean quantity
$g_E^{qq} (\theta;T;t)$, with $\theta \in (0,\pi)$, by an analytic
continuation in the angular variables and in the IR cutoff
\cite{Meggiolaro97,Meggiolaro98,Meggiolaro02}:
\ba
g_E^{qq} (\theta;T;t) &=& g_M^{qq} (\chi \to i\theta;T \to -iT;t) ,
\nonumber \\
g_M^{qq} (\chi;T;t) &=& g_E^{qq} (\theta \to -i\chi;T \to iT;t) .
\label{original}
\ea
This result is derived under certain hypotheses of analyticity in the angular
variables and in the IR cutoff $T$. In particular, one makes
the assumption that the function $g_M$, as a function of the {\it complex}
variable $\chi$, can be {\it analytically extended} from the positive
real axis $(\re\chi > 0, \im\chi = 0)$ to a domain ${\cal D}_M$ which also
includes the imaginary segment $(\re\chi = 0, 0 < \im\chi < \pi)$;
and, therefore, the function $g_E$, as a function of the {\it complex}
variable $\theta$, can be {\it analytically extended} from the real segment
$(0 < \re\theta < \pi, \im\theta = 0)$ to a domain ${\cal D}_E =
\{ \theta \in \mathbb{C} ~|~ i\theta \in {\cal D}_M \}$, which also includes
the negative imaginary axis $(\re\theta = 0, \im\theta < 0)$.
(The validity of this assumption is confirmed by explicit calculations in
perturbation theory \cite{Meggiolaro97}, as we shall also see in section 4.)
To avoid possible confusions, we shall denote with $\overline{g}_M$ and
$\overline{g}_E$ such analytic extensions.\footnote{Of course, if the
domains ${\cal D}_M$ and ${\cal D}_E$ for the analytic extensions
$\overline{g}_M$ and $\overline{g}_E$ include portions of the respective real
axes which are larger than, respectively, the positive real axis
$(\re\chi > 0, \im\chi = 0)$ and the real segment $(0 < \re\theta < \pi,
\im\theta = 0)$, one in general has that $\overline{g}_M \neq g_M$ for
$\chi\in\mathbb{R}^-$ and $\overline{g}_E \neq g_E$ for $\theta\in\mathbb{R},
\theta \not\in (0,\pi)$. For example, while the functions $g_M$ and $g_E$
satisfy the symmetry properties (\ref{propM}) and (\ref{propE1}),
(\ref{propE2}), their analytic extensions $\overline{g}_M$ and
$\overline{g}_E$ might well not satisfy these properties.
We shall see explicit examples of this in section 4.}
Eq. (\ref{original}) is then intended to be valid for every
$\chi \in {\cal D}_M$ (i.e., for every $\theta \in {\cal D}_E$):
\ba
\overline{g}_E^{qq} (\theta;T;t) &=&
\overline{g}_M^{qq} (i\theta;-iT;t) , \qquad \forall\theta\in {\cal D}_E ;
\nonumber \\
\overline{g}_M^{qq} (\chi;T;t) &=&
\overline{g}_E^{qq} (-i\chi;iT;t) , \qquad \forall\chi\in {\cal D}_M .
\label{original-bis}
\ea
This result is valid both for Abelian and non--Abelian gauge theories.
We stress the fact that the {\it regularized} quantities
$g_M(\chi;T;t)$ and $g_E(\theta;T;t)$, while being finite at any given
value of $T$, are divergent in the limit $T \to \infty$
(even if in some cases this IR divercence can be factorized out and one
thus ends up with an IR--finite {\it physical} quantity).

Differently from the parton--parton scattering amplitudes, which are known to
be affected by IR divergences, the elastic scattering amplitude of
two colourless states in gauge theories, e.g., two $q\bar{q}$ meson states,
is expected to be an IR--finite physical quantity \cite{BL}.
It was shown in Refs. \cite{DFK,Nachtmann97,BN} (for a review, see Refs.
\cite{Dosch,pomeron-book}) that the high--energy meson--meson elastic
scattering amplitude can be approximately reconstructed by first evaluating,
in the eikonal approximation, the elastic scattering amplitude of two
$q\bar{q}$ pairs (usually called ``{\it dipoles}''), of given transverse
sizes $\vec{R}_{1\perp}$ and $\vec{R}_{2\perp}$ respectively, and then
averaging this amplitude over all possible values of $\vec{R}_{1\perp}$ and
$\vec{R}_{2\perp}$ with two proper squared wave functions
$|\psi_1 (\vec{R}_{1\perp})|^2$ and $|\psi_2 (\vec{R}_{2\perp})|^2$,
describing the two interacting mesons.\footnote{Here and in what follows we
take, for simplicity, the longitudinal--momentum fractions $f_1$ and $f_2$ of
the two quarks in the two dipoles (and, therefore, also the
longitudinal--momentum fractions $1-f_1$ and $1-f_2$ of the two antiquarks
in the two dipoles) to be fixed to $1/2$: this is known to be a good
approximation for hadron--hadron interactions (see Refs.
\cite{Dosch,pomeron-book} and references therein). However, the dipendence
on the longitudinal--momentum fractions $f_1$ and $f_2$ could be easily
implemented in the hadron wave functions $\psi_1 (\vec{R}_{1\perp},f_1)$ and
$\psi_2 (\vec{R}_{2\perp},f_2)$ and in the loop--loop correlator itself (see
again Refs. \cite{Dosch,pomeron-book} and references therein for more details),
without altering any relevant formula or conclusion in our paper.}
(For the treatment of baryons, a similar, but, of course, more involved,
picture can be adopted, using a genuine three--body configuration or,
alternatively and even more simply, a quark--diquark configuration: we refer
the interested reader to the above--mentioned original references
\cite{Dosch,pomeron-book,DFK,Nachtmann97,BN}.)

The high--energy elastic scattering amplitude of two {\it dipoles} is
governed by the (properly normalized) correlation function of two Wilson loops
${\cal W}_1$ and ${\cal W}_2$, which follow the classical straight lines for
quark (antiquark) trajectories:
\be
{\cal M}_{(ll)} (s,t;\vec{R}_{1\perp},\vec{R}_{2\perp}) \equiv
-i~2s \displaystyle\int d^2 \vec{z}_\perp
e^{i \vec{q}_\perp \cdot \vec{z}_\perp}
\left[ {\langle {\cal W}_1 {\cal W}_2 \rangle \over
\langle {\cal W}_1 \rangle \langle {\cal W}_2 \rangle} -1 \right] ,
\label{scatt-loop}
\ee
where $s$ and $t = -|\vec{q}_\perp|^2$ ($\vec{q}_\perp$ being the tranferred
momentum) are the usual Mandelstam variables.
More explicitly the Wilson loops ${\cal W}_1$ and ${\cal W}_2$ are so defined:
\ba
{\cal W}^{(T)}_1 &\equiv&
{1 \over N_c} \Tr \left\{ {\cal P} \exp
\left[ -ig \displaystyle\oint_{{\cal C}_1} A_\mu(x) dx^\mu \right] \right\} ,
\nonumber \\
{\cal W}^{(T)}_2 &\equiv&
{1 \over N_c} \Tr \left\{ {\cal P} \exp
\left[ -ig \displaystyle\oint_{{\cal C}_2} A_\mu(x) dx^\mu \right] \right\} ,
\label{QCDloops}
\ea
where ${\cal P}$ denotes the ``{\it path ordering}'' along the given path
${\cal C}$; ${\cal C}_1$ and ${\cal C}_2$ are two rectangular paths which
follow the classical straight lines for quark [$X_{(+)}(\tau)$, forward in
proper time $\tau$] and antiquark [$X_{(-)}(\tau)$, backward in $\tau$]
trajectories, i.e.,
\ba
{\cal C}_1 &\to&
X_{(\pm 1)}^\mu(\tau) = z^\mu + {p_1^\mu \over m} \tau
\pm {R_1^\mu \over 2} , \nonumber \\
{\cal C}_2 &\to&
X_{(\pm 2)}^\mu(\tau) = {p_2^\mu \over m} \tau \pm {R_2^\mu \over 2} ,
\label{traj}
\ea
and are closed by straight--line paths at proper times $\tau = \pm T$, where
$T$ plays the role of an IR cutoff, which must be removed in the end
($T \to \infty$).
Here $p_1$ and $p_2$ are the four--momenta of the two dipoles with mass $m$,
moving with speed $V$ and $-V$ along, for example, the $x^1$--direction.
Their expression is given by Eq. (\ref{p1p2}), where $\chi = 2~{\rm arctanh} V$
is the hyperbolic angle between the two trajectories $(+1)$ and $(+2)$.
Moreover, $R_1 = (0,0,\vec{R}_{1\perp})$, $R_2 = (0,0,\vec{R}_{2\perp})$
and $z = (0,0,\vec{z}_\perp)$, where $\vec{z}_\perp = (z^2,z^3)$ is the
impact--parameter distance between the two loops in the transverse plane.

In the Euclidean theory, one considers the correlation function of two
Euclidean Wilson loops $\widetilde{\cal W}_1$ and $\widetilde{\cal W}_2$
running along two rectangular paths $\widetilde{\cal C}_1$ and
$\widetilde{\cal C}_2$ which follow the following
straight--line trajectories:
\ba
\widetilde{\cal C}_1 &\to&
X^{(\pm 1)}_{E\mu}(\tau) = z_{E\mu} + {p_{1E\mu} \over m}
\tau \pm {R_{1E\mu} \over 2} , \nonumber \\
\widetilde{\cal C}_2 &\to&
X^{(\pm 2)}_{E\mu}(\tau) = {p_{2E\mu} \over m} \tau
\pm {R_{2E\mu} \over 2} ,
\label{trajE}
\ea
and are closed by straight--line paths at proper times $\tau = \pm T$. Here
$R_{1E} = (0,\vec{R}_{1\perp},0)$, $R_{2E} = (0,\vec{R}_{2\perp},0)$,
$z_E = (0,\vec{z}_\perp,0)$, and the Euclidean four--vectors $p_{1E}$ and
$p_{2E}$ are defined by Eq. (\ref{p1p2E}), where $\theta$ is the angle formed
by the two trajectories $(+1)$ and $(+2)$ in Euclidean four--space.\\
Let us introduce the following notations for the normalized correlators
$\langle {\cal W}_1 {\cal W}_2 \rangle / \langle {\cal W}_1 \rangle
\langle {\cal W}_2 \rangle$ in the Minkowskian and in the Euclidean theory,
in the presence of a {\it finite} IR cutoff $T$:
\ba
{\cal G}_M(\chi;T;\vec{z}_\perp,\vec{R}_{1\perp},\vec{R}_{2\perp}) &\equiv&
{ \langle {\cal W}^{(T)}_1 {\cal W}^{(T)}_2 \rangle \over
\langle {\cal W}^{(T)}_1 \rangle
\langle {\cal W}^{(T)}_2 \rangle } ,\nonumber \\ 
{\cal G}_E(\theta;T;\vec{z}_\perp,\vec{R}_{1\perp},\vec{R}_{2\perp}) &\equiv&
{ \langle \widetilde{\cal W}^{(T)}_1 \widetilde{\cal W}^{(T)}_2 \rangle_E \over
\langle \widetilde{\cal W}^{(T)}_1 \rangle_E
\langle \widetilde{\cal W}^{(T)}_2 \rangle_E } .
\label{GM-GE}
\ea
As already stated in Ref. \cite{Meggiolaro02}, and formally proved in Ref.
\cite{Meggiolaro05}, the two quantities in Eq. (\ref{GM-GE}) are expected to
be connected by the same analytic continuation in the angular variables and
in the IR cutoff which was already derived in the case of Wilson lines; i.e.,
with analogous hypotheses of analyticity in the angular variables and in
the IR cutoff $T$ and using the same notation already introduced
for the line--line case:
\ba
\overline{\cal G}_E(\theta;T;\vec{z}_\perp,\vec{R}_{1\perp},\vec{R}_{2\perp})
&=& \overline{\cal G}_M
(i\theta;-iT;\vec{z}_\perp,\vec{R}_{1\perp},\vec{R}_{2\perp}) ,
\qquad \forall\theta\in {\cal D}_E ;
\nonumber \\
\overline{\cal G}_M(\chi;T;\vec{z}_\perp,\vec{R}_{1\perp},\vec{R}_{2\perp})
&=& \overline{\cal G}_E
(-i\chi;iT;\vec{z}_\perp,\vec{R}_{1\perp},\vec{R}_{2\perp}) ,
\qquad \forall\chi\in {\cal D}_M .
\label{analytic}
\ea
The analytic continuation (\ref{analytic}) (as the corresponding result for the
line--line case) is an {\it exact}, i.e., {\it nonperturbative} result,
valid both for the Abelian and the non--Abelian case.

As we have said above, the loop--loop correlation functions (\ref{GM-GE}),
both in the Minkowskian and in the Euclidean theory, are expected to be
IR--{\it finite} quantities, i.e., to have finite limits when $T \to \infty$,
differently from what happens in the case of Wilson lines.
One can then define the following loop--loop correlation functions
with the IR cutoff removed:
\ba
{\cal C}_M(\chi;\vec{z}_\perp,\vec{R}_{1\perp},\vec{R}_{2\perp}) &\equiv&
\displaystyle\lim_{T \to \infty} \left[
{\cal G}_M(\chi;T;\vec{z}_\perp,\vec{R}_{1\perp},\vec{R}_{2\perp})
- 1 \right] , \nonumber \\
{\cal C}_E(\theta;\vec{z}_\perp,\vec{R}_{1\perp},\vec{R}_{2\perp})
&\equiv& \displaystyle\lim_{T \to \infty} \left[
{\cal G}_E(\theta;T;\vec{z}_\perp,\vec{R}_{1\perp},\vec{R}_{2\perp})
- 1 \right] .
\label{C12}
\ea
It has been proved in Ref. \cite{Meggiolaro05} that, under certain analyticity
conditions in the {\it complex} variable $T$ [conditions which are also
sufficient to make the relations (\ref{analytic}) meaningful], the two
quantities (\ref{C12}), obtained {\it after} the removal of the IR cutoff
($T \to \infty$), are still connected by the usual analytic continuation in
the angular variables only:
\ba
\overline{\cal C}_E(\theta;\vec{z}_\perp,\vec{R}_{1\perp},\vec{R}_{2\perp}) &=&
\overline{\cal C}_M(i\theta;\vec{z}_\perp,\vec{R}_{1\perp},\vec{R}_{2\perp}) ,
\qquad \forall\theta\in {\cal D}_E ;
\nonumber \\
\overline{\cal C}_M(\chi;\vec{z}_\perp,\vec{R}_{1\perp},\vec{R}_{2\perp}) &=&
\overline{\cal C}_E(-i\chi;\vec{z}_\perp,\vec{R}_{1\perp},\vec{R}_{2\perp}) ,
\qquad \forall\chi\in {\cal D}_M .
\label{final}
\ea
This is a highly non--trivial result, whose general validity is discussed
in Ref. \cite{Meggiolaro05}.

As said in Ref. \cite{Meggiolaro05},
if ${\cal G}_M$ and ${\cal G}_E$, considered as functions of the
{\it complex} variable $T$, have in $T=\infty$ an ``eliminable {\it isolated}
singular point'' [i.e., they are analytic functions of $T$ in the {\it complex}
region $|T| > R$, for some $R\in\mathbb{R}^+$, and the {\it finite} limits
(\ref{C12}) exist when letting the {\it complex} variable $T \to \infty$],
then, of course, the analytic continuation (\ref{final}) immediately derives
from Eq. (\ref{analytic}) (with $|T| > R$), when letting
$T \to +\infty$.\footnote{For example, if ${\cal G}_M$ and ${\cal G}_E$ are
analytic functions of $T$ in the {\it complex} region $|T| > R$, for some
$R\in\mathbb{R}^+$, and they are bounded at large $T$, i.e.,
$\exists B_{M,E}\in\mathbb{R}^+$ such that $|{\cal G}_{M,E}(T)| < B_{M,E}$
for $|T| > R$, then $T=\infty$ is an ``eliminable singular point''
for both of them.}
But the same result (\ref{final}) can also be derived under different
conditions. For example, let us assume that ${\cal G}_E$ is a bounded
analytic function of $T$ in the sector $0 \le \arg T \le {\pi \over 2}$,
with finite limits along the two straight lines on the border of the sector:
${\cal G}_E \to G_{E1}$, for $({\re}T \to +\infty,~{\im}T = 0)$, and
${\cal G}_E \to G_{E2}$, for $({\re}T = 0,~{\im}T \to +\infty)$.
And, similarly, let us assume that ${\cal G}_M$ is a bounded
analytic function of $T$ in the sector $-{\pi \over 2} \le \arg T \le 0$,
with finite limits along the two straight lines on the border of the sector:
${\cal G}_M \to G_{M1}$, for $({\re}T \to +\infty,~{\im}T = 0)$, and
${\cal G}_M \to G_{M2}$, for $({\re}T = 0,~{\im}T \to -\infty)$.
We can then apply the ``Phragm\'en--Lindel\"of theorem'' (see, e.g., Theorem
5.64 in Ref. \cite{PLT}) to state that $G_{E2} = G_{E1}$ and
$G_{M2} = G_{M1}$. Therefore, also in this case, the analytic continuation
(\ref{final}) immediately derives from Eq. (\ref{analytic}) when
$T \to \infty$.

\newsection{Analyticity and crossing symmetry}

\noindent
In this section we will show how the analytic--continuation relations from
the Minkowskian to the Euclidean theory lead to a nice geometrical
interpretation of the so--called {\it crossing symmetry} between the
quark--quark and quark--antiquark scattering amplitudes (and also between
dipole--dipole scattering amplitudes) in the eikonal approximation.

In such an approximation the scattering amplitudes
factorize in a product of Kronecker's deltas in the
spin variables, expressing spin conservation at high energies, and in
a term that is  essentially the (normalized)
correlator of two Wilson lines in the appropriate representation.
According to the results found in \cite{Nachtmann97} and \cite{Meggiolaro01},
changing from a quark to an antiquark just corresponds, in our formalism,
to substitute the corresponding Wilson line (in the fundamental representation)
with its complex conjugate (i.e., the Wilson line in the complex conjugate
representation, $T'_a = -T^*_a$).
Therefore, the eikonal amplitude for the soft elastic
scattering of a quark $q$ and an antiquark $\bar{q}$ with given spin and
colour quantum numbers,
\be
q(p_1,\alpha,i) + \bar{q}(p_2,\beta,j) \to
q(p_1'\simeq p_1,\alpha',i') + \bar{q}(p_2'\simeq p_2, \beta',j') ,
\ee
where the particles four--momenta $p_1' \simeq p_1$ and $p_2' \simeq p_2$
are defined in Eq. (\ref{p1p2}), is given by the formula:
\be
\mathcal{M}^{q\bar{q}}(p_1,p_2;t)_{i'i;j'j}^{\alpha'\alpha;\beta'\beta}
\mathop{\sim}_{s \to \infty}
-i~ 2s~ \delta_{\alpha'\alpha}\delta_{\beta'\beta}~
g_M^{q\bar{q}}(p_1,p_2;T\to\infty;t)_{i'i;j'j} ,
\ee
where the correlator $g_M^{q\bar{q}}(p_1,p_2;T;t)_{i'i;j'j}$
is defined as:\footnote{The Wilson--line renormalization constants in
the complex conjugate and in the fundamental representations are equal
because of the invariance of the functional integral under charge conjugation
of the gluon fields, i.e., $A_{\mu} \to A_{\mu}' = -A_{\mu}^T = -A_{\mu}^*$:
$\frac{1}{N_c}\langle\textrm{Tr}[W^{(T)*}_p(\vec{0}_\perp)]\rangle
= \frac{1}{N_c}\langle\textrm{Tr}[W^{(T)}_p(\vec{0}_\perp)]\rangle = Z_M(T).$}
\be
\label{eq:anti1}
g_M^{q\bar{q}}(p_1,p_2;T;t)_{i'i;j'j} = \frac{1}{[Z_M(T)]^2}
\int d^2\vec{z}_{\perp}e^{i\vec{q}_{\perp}\cdot\vec{z}_{\perp}}
\langle [W_{p_1}^{(T)}(\vec{z}_\perp) - \mathbb{I}]_{i'i}
[W_{p_2}^{(T)*}(\vec{0}_\perp) - \mathbb{I}]_{j'j}\rangle .
\ee
Crossing symmetry relates the amplitude of this
process to the amplitude of the {\it ``crossed''} process, defined as:
\be
q(p_1,\alpha,i) + q(-p_2'\simeq -p_2, \beta',j') \to
q(p_1'\simeq p_1,\alpha',i') + q(-p_2,\beta,j).
\ee
Using the fact that the generators $T_a$ are hermitian and the
variables $A_{\mu}^a$ are real, the complex conjugate Wilson line
$W_{p_2}^{(T)*}(\vec{0}_\perp)$ corresponding to the antiquark can also
be re--written as follows:
\ba
\lefteqn{
\left[W_{p_2}^{(T)*}(\vec{0}_\perp)\right]_{j'j} =
\left[{\cal T}\exp\left(ig\int_{-T}^{+T}A^*_{\mu}(\frac{p_2}{m}\tau)
\frac{p_2^{\mu}}{m} d\tau\right)\right]_{j'j} }
\nonumber \\
& & =\sum_{n=0}^{\infty}\int_{-T}^{+T} d\tau_1\ldots
\int_{-T}^{+T}d\tau_n \theta(\tau_1 - \tau_2)
\ldots\theta(\tau_{n-1}-\tau_n)
\nonumber \\
& & \times\left\{\left[igA^*_{\mu_1}(\frac{p_2}{m}\tau_1)\frac{p_2^{\mu_1}}{m}
\right]\ldots
\left[igA^*_{\mu_n}(\frac{p_2}{m}\tau_n)\frac{p_2^{\mu_n}}{m}
\right]\right\}_{j'j}
\nonumber \\
& & =\sum_{n=0}^{\infty}\int_{-T}^{+T} d\tau_1\ldots
\int_{-T}^{+T}d\tau_n \theta(\tau_1 - \tau_2)\ldots\theta(\tau_{n-1}-\tau_n)
\nonumber \\
& & \times\left\{\left[igA_{\mu_n}(\frac{p_2}{m}\tau_n)\frac{p_2^{\mu_n}}{m}
\right]\ldots\left[igA_{\mu_1}(\frac{p_2}{m}\tau_1)
\frac{p_2^{\mu_1}}{m}\right]\right\}_{jj'}
\nonumber \\
& & =\left[\overline{\cal T}
\exp\left(ig\int_{-T}^{+T}A_{\mu}(\frac{p_2}{m}\tau)
\frac{p_2^{\mu}}{m} d\tau\right)\right]_{jj'} = 
\left[W_{p_2}^{(T)}(\vec{0}_\perp)^{\dag}\right]_{jj'} ,
\label{W2*}
\ea
where $\overline{\cal T}\exp(\ldots)$ is the ``anti ${\cal T}$--ordered''
exponential. Replacing the integration variables
$\tau_i = -\tau_i'$ in the last expression we immediately get:
\ba
\lefteqn{
\left[W_{p_2}^{(T)*}(\vec{0}_\perp)\right]_{j'j} =
\left[W_{p_2}^{(T)}(\vec{0}_\perp)^{\dag}\right]_{jj'} }
\nonumber \\
& & =\sum_{n=0}^{\infty}\int_{-T}^{+T} d\tau_n'\ldots
\int_{-T}^{+T}d\tau_1' \theta(\tau_n' - 
\tau_{n-1}')\ldots\theta(\tau_2'-\tau_1')
\nonumber \\
& & \times\left\{\left[-igA_{\mu_n}(-\frac{p_2}{m}\tau_n')
\left(-\frac{p_2^{\mu_n}}{m}\right)\right]\ldots
\left[-igA_{\mu_1}(-\frac{p_2}{m}\tau_1')\left(-\frac{p_2^{\mu_1}}{m}
\right)\right]\right\}_{jj'}
\nonumber \\
& & \equiv \left[W_{-p_2}^{(T)}(\vec{0}_\perp)\right]_{jj'} .
\ea
Summarizing:
\be 
\label{eq:croce1}
\left[W_{p_2}^{(T)*}(\vec{0}_\perp)\right]_{j'j} =
\left[W_{p_2}^{(T)}(\vec{0}_\perp)^{\dag}\right]_{jj'}=
\left[W_{-p_2}^{(T)}(\vec{0}_\perp)\right]_{jj'}.
\ee
We can now write the correlator $g_M^{q\bar{q}}$ in the form:
\be
\label{eq:anti2}
g_M^{q\bar{q}}(p_1,p_2;T;t)_{i'i;j'j} = \frac{1}{[Z_M(T)]^2}
\int d^2\vec{z}_{\perp}e^{i\vec{q}_{\perp}\cdot\vec{z}_{\perp}}
\langle [W_{p_1}^{(T)}(\vec{z}_\perp) - \mathbb{I}]_{i'i}
[W_{-p_2}^{(T)}(\vec{0}_\perp) - \mathbb{I}]_{jj'}\rangle ;
\ee
that is, reminding the definition of the quark--quark correlator:
\be
\label{eq:croce2}
g_M^{q\bar{q}}(p_1,p_2;T;t)_{i'i;j'j} = g_M^{qq}(p_1,-p_2;T;t)_{i'i;jj'} .
\ee
This relation is the direct expression of crossing
symmetry, once we have formulated the appropriate analyticity conditions 
on $g_M^{qq}$ as a function of the four--momenta making
the right--hand side meaningful, and is valid for every value of the
IR cutoff $T$.

We want to give now a ``geometrical'' interpretation
of this relation, expressing it in terms of the hyperbolic angle $\chi$ 
between the four--momenta $p_1$ and $p_2$; using this
interpretation we will be able to discuss in details the analyticity 
hypotheses on $g_M^{qq}$ and $g_M^{q\bar{q}}$ that make
the relation (\ref{eq:croce2}) meaningful. 

We shall denote the left--hand side of (\ref{eq:croce2}) (with a slight abuse
of notation) also as $g_M^{q\bar{q}}(\chi;T;t)_{i'i;j'j}$;
in the right--hand side we have instead the
function $g_M^{qq}(p_1,\tilde{p}_2;T;t)_{i'i;jj'}$ calculated at four--momenta
$p_1$ and $\tilde{p}_2 = -p_2$; the substitution of $p_2$ with the
(unphysical) four--momentum $\tilde{p}_2$ corresponds to the substitution
$\cosh\chi\to -\cosh\chi$. To determine unambiguously which complex
values of $\chi$ this substitution corresponds to, we will make use of 
the analytic--continuation relation between the Minkowskian and the
Euclidean theory and of the $O(4)$ symmetry of the latter. 

The relation (\ref{eq:croce1}) is evidently valid also
for Euclidean Wilson lines, i.e.,
\be
\label{eq:croce3}
\left[\widetilde{W}_{p_{2E}}^{(T)*}(\vec{0}_\perp)\right]_{j'j} =
\left[\widetilde{W}_{p_{2E}}^{(T)}(\vec{0}_\perp)^{\dag}\right]_{jj'}=
\left[\widetilde{W}_{-p_{2E}}^{(T)}(\vec{0}_\perp)\right]_{jj'} ,
\ee
and so relation (\ref{eq:croce2}) is extended to the
Euclidean case:
\be
\label{eq:croce4}
g_E^{q\bar{q}}(p_{1E},p_{2E};T;t)_{i'i;j'j} =
g_E^{qq}(p_{1E},-p_{2E};T;t)_{i'i;jj'} ,
\ee
where the Euclidean four--momenta $p_{1E}$ and $p_{2E}$ are given by
Eq. (\ref{p1p2E}).
In our notation the left--hand side of (\ref{eq:croce4}) is denoted as
$g_E^{q\bar{q}}(\theta;T;t)_{i'i;j'j}$, where $\theta$ is the angle
between the Euclidean four--momenta $p_{1E}$ and $p_{2E}$.
The right--hand side can be written as $g_E^{qq}(\pi+\theta;T;t)_{i'i;jj'}$,
using the invariance under the $O(4)$  $90^{\circ}$
clockwise ``rotation'' in the $(x_{E1},x_{E4})$ plane:
\be
\label{eq:trasfo}
x_E \to x_E' = {\cal R}_{3}x_E, \qquad {\cal R}_{3} = \left( 
\begin{array}{cccc}
0 & 0 & 0 & 1\\
0 & 1 & 0 & 0\\
0 & 0 & 1 & 0\\
-1 & 0 & 0 & 0
\end{array}\right) ,
\ee
and also, using the relation (\ref{propE2}),
as $g_E^{qq}(\pi-\theta;T;t)_{i'i;jj'}$. In this way
relation (\ref{eq:croce4}) takes the form (see Fig. 2):
\be  
\label{eq:croce5}
g_E^{q\bar{q}}(\theta;T;t)_{i'i;j'j} = g_E^{qq}(\pi -\theta;T;t)_{i'i;jj'},
\qquad \forall\theta\in\mathbb{R}.
\ee
For $\theta\in (0,\pi)$ the two functions
$g_E^{q\bar{q}}$ and $g_E^{qq}$ are calculated in
points belonging to the respective analyticity domains.
Suppose now that the relation (\ref{eq:croce5}) can be
analytically extended to complex values of $\theta$ in
a common analyticity domain ${\cal D}_E$ for $\overline{g}_E^{qq}$ and
$\overline{g}_E^{q\bar{q}}$ (and for every value of the {\it complex} variable
$T$ in an appropriate analyticity domain). This domain must then have the
property that, if $\theta \in {\cal D}_E$, then also
$\pi - \theta \in {\cal D}_E$, i.e., it has to be symmetric with
respect to the point $\theta_0 = ({\re}\theta_0 =
\pi/2, {\im}\theta_0 = 0)$, and it has to include the segment
$(0<{\re}\theta < \pi, {\im}\theta = 0)$, the  negative imaginary
axis $({\re}\theta =0, {\im}\theta < 0)$
and the semiaxis $({\re}\theta = \pi, {\im}\theta > 0)$:
it is schematically shown in Fig. 3.\\
Using the notation previously introduced, we write:
\be
\label{eq:croce6}
\overline{g}_E^{q\bar{q}}(\theta;T;t)_{i'i;j'j} =
\overline{g}_E^{qq}(\pi -\theta;T;t)_{i'i;jj'},
\qquad \forall\theta\in\mathcal{D}_E.
\ee
Now, using repeatedly the analytic--continuation relations
(\ref{original-bis}), we get the following relation between
the Minkowskian correlators:
\ba
\lefteqn{
\overline{g}_M^{q\bar{q}}(\chi;T;t)_{i'i;j'j} =
\overline{g}_E^{q\bar{q}}(-i\chi;iT;t)_{i'i;j'j} =
\overline{g}_E^{qq}(\pi + i\chi;iT;t)_{i'i;jj'} }
\nonumber \\ 
& & =\overline{g}_E^{qq}\left(-i\left(i\pi -\chi\right);iT;t\right)_{i'i;jj'}
=\overline{g}_M^{qq}(i\pi - \chi;T;t)_{i'i;jj'},
\qquad \forall\chi\in\mathcal{D}_M ,
\label{eq:croce7}
\ea
where $\mathcal{D}_M = \{\chi\in\mathbb{C}|-i\chi\in\mathcal{D}_E\}$ is the
common analyticity domain of $\overline{g}_M^{qq}$ and
$\overline{g}_M^{q\bar{q}}$, with the property that,
if $\chi\in\mathcal{D}_M$, then also $i\pi-\chi\in\mathcal{D}_M$,
i.e., it is symmetric with respect to the point $\chi_0 = ({\re}\chi_0 = 0,\,
{\im}\chi_0 = \pi/2)$, and it includes the real positive axis,
$({\re}\chi >0,\, {\im}\chi=0)$, the imaginary segment
$({\re}\chi =0,\, 0< {\im}\chi<\pi)$
and the semiaxis $({\re}\chi <0,\, {\im}\chi =\pi)$:
it is schematically shown in Fig. 4.\\
In particular, for $\chi\in\mathbb{R}^+$ we have:
\be
\label{eq:croce8}
g_M^{q\bar{q}}(\chi;T;t)_{i'i;j'j} =
\overline{g}_M^{qq}(i\pi - \chi;T;t)_{i'i;jj'},
\qquad \forall\chi\in\mathbb{R}^+.
\ee
This is the ``geometrical transcription'' in terms of the angular variable
$\chi$ of relation (\ref{eq:croce2}), and states that the quark--antiquark
correlator can be obtained from the quark--quark one by the
analytic--continuation $\chi\to i\pi-\chi$ in the hyperbolic angle and by the
colour index exchange $j\leftrightarrow j'$.\footnote{In Ref. \cite{JP2} the
crossing--symmetry relation for line--line correlators was instead
identified with $\chi \to \chi - i\pi$. The correct relation
$\chi \to i\pi - \chi$ has been {\it guessed}, but not properly justified,
in Ref. \cite{Korchemsky}, apparently only on the basis of the
correspondence rule $\cosh\chi \to -\cosh\chi$, which however, as we have
already said, cannot unambiguously fix the correspondence rule for the
hyperbolic angle $\chi$ alone.}

Reminding the relation (\ref{s-chi}) between the Mandelstam variable
$s$ and the hyperbolic angle $\chi$,
we see that the substitution $\chi\to i\pi-\chi$ corresponds,
taking the limit $\chi\to\infty$, to the substitution
\be
s \to e^{-i\pi}s ,
\ee
while the Mandelstam variable $t$ doesn't change going
over to the crossed channel. This is in agreement with
what we expect from crossing symmetry: the exchange $p_2
\leftrightarrow -p_2'$ implies the exchange
$s = (p_1+p_2)^2 \leftrightarrow u = (p_1-p_2')^2$,
while $t = (p_1-p_1')^2$ remains unchanged; moreover, in our limit,
because of the relation $s+t+u=4m^2$, we have $u\simeq -s$.

In a perfectly analogous way we can obtain a crossing--symmetry
relation for loop--loop correlators. Let us consider a certain Wilson loop
\be
\mathcal{W}_p^{(T)}(\vec{b}_{\perp},\vec{R}_{\perp}) =
\frac{1}{N_c}{\rm Tr}\left\{\mathcal{P}\exp\left[     
-ig\oint_{\mathcal{C}(p,b,R)} A_{\mu}(x)dx^{\mu}
\right]\right\} ,
\ee
defined on the rectangular path $\mathcal{C}(p,b,R)$,
consisting of the straight--line trajectories [$b = (0,0,\vec{b}_{\perp})$, 
$R = (0,0,\vec{R}_{\perp})$]
\be
X^{\mu}_{(\pm)}(\tau) = b^{\mu} +
\frac{p^{\mu}}{m}\tau \pm\frac{R^{\mu}}{2}
\ee
of the quark [$X^{\mu}_{(+)}(\tau)$, with $\tau$ going from $-T$ to $+T$]
and of the antiquark
[$X^{\mu}_{(-)}(\tau)$, with $\tau$ going from $+T$ to $-T$], joined by
straight--line paths at $\tau = \pm T$ ({\it ``links''}), so making the loop
a gauge invariant operator. Let us define the corresponding {\it antiloop}
$\overline{\mathcal{W}}$ by exchanging the quark and the antiquark
trajectories (and reversing the links direction in order to preserve
gauge invariance). Clearly this corresponds to reverse the direction
of the path of the initial loop $\mathcal{W}$, i.e., to make the
substitution $p \to -p$:
\be
\overline{\mathcal{W}}_p^{(T)}(\vec{b}_\perp,\vec{R}_\perp) =
\frac{1}{N_c}{\rm Tr}\left\{\mathcal{P}\exp\left[
-ig\oint_{\overline{\mathcal{C}}(p,b,R)} A_{\mu}(x)dx^{\mu}
\right]\right\} ,
\ee
where:
\be
\overline{\mathcal{C}}(p,b,R) = \mathcal{C}(-p,b,R).
\ee
Evidently, the transition from a loop to the
corresponding antiloop can also be made by keeping $p$ fixed
and substituting $R\to -R$. Consequently:
\be
\overline{\mathcal{C}}(p,b,R) = \mathcal{C}(-p,b,R) =
\mathcal{C}(p,b,-R) ,
\ee
and:
\be
\label{eq:croce9}
\overline{\mathcal{W}}_p^{(T)}(\vec{b}_\perp,\vec{R}_\perp) =
\mathcal{W}_{-p}^{(T)}(\vec{b}_\perp,\vec{R}_\perp) =
\mathcal{W}_{p}^{(T)}(\vec{b}_\perp,-\vec{R}_\perp).
\ee
Let us define the loop--antiloop correlator
$\mathcal{G}_M^{(l\bar{l})}$ substituting in the
loop--loop correlator $\mathcal{G}_M$ 
the loop $\mathcal{W}_2$ with the corresponding
antiloop:\footnote{Also in this case the charge--conjugation invariance
(or, more simply, the rotation invariance) imposes that the vacuum
expectation values of the loop and the antiloop are equal:
$\langle \overline{\mathcal{W}} \rangle = \langle \mathcal{W} \rangle.$}
\be
\label{eq:lupo2}
\mathcal{G}_M^{(l\bar{l})}(\chi;T;\vec{z}_{\perp},
\vec{R}_{1\perp},\vec{R}_{2\perp}) = 
\frac{\langle \mathcal{W}^{(T)}_1
\overline{\mathcal{W}}^{(T)}_2 \rangle}{\langle
\mathcal{W}^{(T)}_1\rangle\langle 
\overline{\mathcal{W}}^{(T)}_2 \rangle}.
\ee
Going on as we have done in the line--line case,
we immediately verify that the first equality in
(\ref{eq:croce9}) leads to the crossing--symmetry relation:
\be
\label{eq:croce10}
\mathcal{G}_M^{(l\bar{l})}(\chi;T;\vec{z}_{\perp},\vec{R}_{1\perp},
\vec{R}_{2\perp}) =\overline{\mathcal{G}}_M(i\pi-\chi;T;\vec{z}_{\perp},
\vec{R}_{1\perp},\vec{R}_{2\perp}),
\qquad \forall\chi\in\mathbb{R}^+ .
\ee
As before, it is derived from the Euclidean space
relation obtained from the Euclidean version of
(\ref{eq:croce9}), i.e.,
\be
\label{eq:croce11}
\mathcal{G}_E^{(l\bar{l})}(\theta;T;\vec{z}_{\perp},\vec{R}_{1\perp},
\vec{R}_{2\perp}) =\mathcal{G}_E(\pi-\theta;T;\vec{z}_{\perp},
\vec{R}_{1\perp},\vec{R}_{2\perp}) ,
\qquad \forall\theta\in\mathbb{R} ,
\ee
with appropriate analyticity hypotheses on
$\mathcal{G}_E$ as a function of the angular variable
$\theta$ (or on $\mathcal{G}_M$ as a function of the angular variable
$\chi$), completely analogous to the hypotheses made in the line--line case.
Moreover, the second equality in (\ref{eq:croce9}) implies that:
\be
\label{eq:croce12}
\mathcal{G}_M^{(l\bar{l})}(\chi;T;\vec{z}_{\perp},\vec{R}_{1\perp},
\vec{R}_{2\perp}) =\mathcal{G}_M(\chi;T;\vec{z}_{\perp},\vec{R}_{1\perp},
-\vec{R}_{2\perp});
\ee
and, in the Euclidean case:
\be
\label{eq:croce13}
\mathcal{G}_E^{(l\bar{l})}(\theta;T;\vec{z}_{\perp},\vec{R}_{1\perp},
\vec{R}_{2\perp}) = \mathcal{G}_E(\theta;T;\vec{z}_{\perp},\vec{R}_{1\perp},
-\vec{R}_{2\perp}).
\ee
These two relations, together with the relations (\ref{eq:croce10}) and
(\ref{eq:croce11}) found above, allow us to deduce non trivial properties
of the Minkowskian correlator $\mathcal{G}_M$ under the exchange
$\chi\to i\pi -\chi$ and of the Euclidean correlator $\mathcal{G}_E$
under the exchange $\theta\to \pi -\theta$. In the Minkowskian case:
\ba
\lefteqn{
\overline{\mathcal{G}}_M(i\pi-\chi;T;\vec{z}_{\perp},\vec{R}_{1\perp},
\vec{R}_{2\perp}) }
\nonumber \\
& & =\mathcal{G}_M(\chi;T;\vec{z}_{\perp},\vec{R}_{1\perp},-\vec{R}_{2\perp})
=\mathcal{G}_M(\chi;T;\vec{z}_{\perp},-\vec{R}_{1\perp},\vec{R}_{2\perp}) ,
\qquad \forall\chi\in\mathbb{R}^+ ;
\label{eq:croce14}
\ea
while, in the Euclidean case:
\ba
\lefteqn{
\mathcal{G}_E(\pi-\theta;T;\vec{z}_{\perp},\vec{R}_{1\perp},
\vec{R}_{2\perp}) }
\nonumber \\
& & =\mathcal{G}_E(\theta;T;\vec{z}_{\perp},\vec{R}_{1\perp},-\vec{R}_{2\perp})
=\mathcal{G}_E(\theta;T;\vec{z}_{\perp},-\vec{R}_{1\perp},\vec{R}_{2\perp}) ,
\qquad \forall\theta\in\mathbb{R} .
\label{eq:croce15}
\ea
[The last two equalities in (\ref{eq:croce14}) and
(\ref{eq:croce15}) are obtained considering the exchange $\mathcal{W}_1 \to
\overline{\mathcal{W}}_1$ instead of $\mathcal{W}_2 \to
\overline{\mathcal{W}}_2$.]
These two relations are valid for every value of the IR cutoff $T$ and so
completely analogous relations also holds for the loop--loop correlation
functions ${\cal C}_M$ and ${\cal C}_E$ with the IR cutoff removed
($T \to \infty$), defined in Eq. (\ref{C12}).

\newsection{Perturbative expansion of the eikonal amplitudes}

\noindent
As the {\it exact} (i.e., nonperturbative) calculation from first principles
of the line--line and loop--loop correlators is beyond our possibilities
(but see also the discussion in section 5), we cannot verify directly if they
satisfy the desired analyticity conditions. A way to study the analytic
structure of such correlators is to use perturbation theory.\\
Perturbation theory is the only calculation technique from first principles
available {\it both} in the Minkowskian {\it and} in the Euclidean theory,
and although the properties of the perturbative series to any given order
do not allow us to get conclusive results, they can however give us some
useful insights about the analytic structure of the {\it real}
(nonperturbative) correlation functions.
Let us start considering the loop--loop correlation functions.

As a pedagogic example to illustrate these considerations, we shall first
consider the simple case of QED, in the so--called {\it quenched}
approximation, where vacuum polarization effects, arising from the presence
of loops of dynamical fermions, are neglected: this amounts to putting the
fermion--matrix determinant equal to $1$, i.e., $\det(Q[A]) = 1$ in
Eq. (\ref{fintM}) and $\det(Q^{(E)}[A^{(E)}]) = 1$ in Eq. (\ref{fintE}).
In such an approximation the functional integrals become simple Gaussian
integrals and therefore the calculation of the normalized loop--loop
correlators (\ref{GM-GE}) can be performed {\it exactly} (i.e.,
nonpertubatively) both in Minkowskian and in Euclidean theory. One finds
\cite{Meggiolaro05} that i) the two quantities ${\cal G}_M$ and ${\cal G}_E$
are indeed connected by the analytic continuation (\ref{analytic}), and ii)
the two quantities are finite in the limit when the IR cutoff $T$ goes to
infinity, the two limits (\ref{C12}) being:
\ba
{\cal C}_M(\chi;\vec{z}_\perp,\vec{R}_{1\perp},\vec{R}_{2\perp}) &=&
\exp \left[ -i 4e^2 |\coth \chi|~
t(\vec{z}_\perp,\vec{R}_{1\perp},\vec{R}_{2\perp}) \right] - 1 ,
\nonumber \\
{\cal C}_E(\theta;\vec{z}_\perp,\vec{R}_{1\perp},\vec{R}_{2\perp}) &=&
\exp \left[ - 4e^2 {\cos\theta \over |\sin\theta|}~
t(\vec{z}_\perp,\vec{R}_{1\perp},\vec{R}_{2\perp}) \right] - 1 ,
\label{QED}
\ea
where the coupling constant is now the electric charge $e$ and
\ba
t(\vec{z}_\perp,\vec{R}_{1\perp},\vec{R}_{2\perp}) &\equiv&
\displaystyle\int {d^2 \vec{k}_\perp \over (2\pi)^2}
{e^{-i\vec{k}_\perp \cdot \vec{z}_\perp} \over \vec{k}_\perp^2}
\sin \left( {\vec{k}_\perp \cdot \vec{R}_{1\perp} \over 2} \right)
\sin \left( {\vec{k}_\perp \cdot \vec{R}_{2\perp} \over 2} \right) \nonumber \\
&=& {1 \over 8\pi} \log \left(
{ |\vec{z}_\perp+{\vec{R}_{1\perp} \over 2}+{\vec{R}_{2\perp} \over 2}|
  |\vec{z}_\perp-{\vec{R}_{1\perp} \over 2}-{\vec{R}_{2\perp} \over 2}| \over
  |\vec{z}_\perp+{\vec{R}_{1\perp} \over 2}-{\vec{R}_{2\perp} \over 2}|
  |\vec{z}_\perp-{\vec{R}_{1\perp} \over 2}+{\vec{R}_{2\perp} \over 2}| }
\right) .
\label{t-function}
\ea
One immediately sees that the analytic extension $\overline{\cal C}_M$ of the
Minkowskian correlator from the positive real axis $\chi\in\mathbb{R}^+$
and the analytic extension $\overline{\cal C}_E$ of the Euclidean correlator
from the real segment $\theta\in(0,\pi)$ are given by:
\ba
\label{eq:overC}
\overline{\cal C}_M(\chi;\vec{z}_\perp,\vec{R}_{1\perp},\vec{R}_{2\perp}) &=&
\exp \left[ -i 4e^2 \coth \chi~
t(\vec{z}_\perp,\vec{R}_{1\perp},\vec{R}_{2\perp}) \right] - 1 ,
\quad \forall\chi\in{\cal D}_M ;
\nonumber \\
\overline{\cal C}_E(\theta;\vec{z}_\perp,\vec{R}_{1\perp},\vec{R}_{2\perp}) &=&
\exp \left[ - 4e^2 \cot \theta~
t(\vec{z}_\perp,\vec{R}_{1\perp},\vec{R}_{2\perp}) \right] - 1 ,
\qquad \forall\theta\in{\cal D}_E .
\ea
The analyticity domain ${\cal D}_M$ of $\overline{\cal C}_M$ in the complex
variable $\chi$ is the entire complex plane with the exception of the points
$ik\pi$, $k\in\mathbb{Z}$, and, equivalently, the analyticity domain
${\cal D}_E$ of $\overline{\cal C}_E$ in the complex variable $\theta$ is
the entire complex plane with the exception of the points $k\pi$,
$k\in\mathbb{Z}$, i.e.,
\ba
{\cal D}_M &=& \{\chi \in \mathbb{C} | \chi \ne ik\pi,~ k \in \mathbb{Z}\};
\nonumber \\
{\cal D}_E &=& \{\theta \in \mathbb{C} | i\theta \in {\cal D}_M\}
= \{\theta \in \mathbb{C} | \theta \ne k\pi,~ k \in \mathbb{Z}\}.
\label{domain}
\ea
These domains have precisely the characteristics, described in the previous
sections, which are sufficient to guarantee both the analytic--continuation
relations (\ref{final}) and the crossing--symmetry relations
(\ref{eq:croce14}) and (\ref{eq:croce15}), with $T \to \infty$.
And these relations are indeed realized by the explicit expressions
(\ref{eq:overC}).
(Also the presence of the singularities for $\chi = ik\pi$, $k\in\mathbb{Z}$,
or, equivalently, for $\theta = k\pi$, $k\in\mathbb{Z}$, are not unexpected
and they are discussed in appendix B.)

As shown in Ref. \cite{Meggiolaro05}, the results (\ref{QED})
can be used to derive the corresponding results in the case of a
non--Abelian gauge theory with $N_c$ colours, up to the order ${\cal O}(g^4)$
in perturbation theory (see also Refs. \cite{BB,KL,LLCM2}):
\ba
{\cal C}_M(\chi;\vec{z}_\perp,\vec{R}_{1\perp},\vec{R}_{2\perp})|_{g^4} &=&
- 2g^4 \left( {N_c^2 - 1 \over N_c^2} \right) \coth^2 \chi~
[t(\vec{z}_\perp,\vec{R}_{1\perp},\vec{R}_{2\perp})]^2 ,
\nonumber \\
{\cal C}_E(\theta;\vec{z}_\perp,\vec{R}_{1\perp},\vec{R}_{2\perp})|_{g^4} &=&
2g^4 \left( {N_c^2 - 1 \over N_c^2} \right) \cot^2 \theta~
[t(\vec{z}_\perp,\vec{R}_{1\perp},\vec{R}_{2\perp})]^2 .
\label{QCD-pert}
\ea
In this case, obviously, these are also the expressions for the analytic
extension $\overline{\cal C}_M$ from the positive real axis
$\chi\in\mathbb{R}^+$ and the analytic extension $\overline{\cal C}_E$
from the real segment $\theta\in(0,\pi)$,
with analyticity domains ${\cal D}_M$ and ${\cal D}_E$
still given by Eq. (\ref{domain}).  Both the analytic--continuation
relations (\ref{final}) and the crossing--symmetry relations
(\ref{eq:croce14}) and (\ref{eq:croce15}), with $T \to \infty$,
are of course trivially satisfied.
(Indeed, the validity of the relation (\ref{final}) for the loop--loop
correlators has been also recently verified in Ref. \cite{BB} by an explicit
calculation up to the order ${\cal O}(g^6)$ in perturbation theory.)

In the case of the line--line correlators we cannot simply remove the
IR cutoff $T$, as we have done in Eq. (\ref{C12}) for the loop--loop case,
since the limits $T \to \infty$ are divergent.
Nevertheless, we can remove the IR cutoff $T$, by letting $T \to \infty$,
provided that {\it another} IR cutoff $\lambda$ has been introduced to
regularize the line--line correlators.
This is exactly what one usually does when one computes the correlators in
perturbation theory by giving a small mass $\lambda$ to the gluons (or
photons) exchanged in each graph. In this way one can define the new
IR--regularized line--line correlators $g_M^{(\lambda)}$ and
$g_E^{(\lambda)}$ by removing the {\it nonperturbative} IR cutoff $T$
($T \to \infty$), while keeping the {\it perturbative} IR cutoff $\lambda$
fixed, i.e.:
\ba
g_M^{(\lambda)}(\chi;t) &\equiv& \displaystyle\lim_{T \to \infty}
g_M^{(\lambda)}(\chi;T;t) , \nonumber \\
g_E^{(\lambda)}(\theta;t) &\equiv& \displaystyle\lim_{T \to \infty}
g_E^{(\lambda)}(\theta;T;t) .
\label{g-lambda}
\ea
Then we can repeat what we have said and done above, at the end of section 2,
for the loop--loop correlators and thus conclude that, under certain analogous
analyticity conditions in the {\it complex} variable $T$ for the two
IR--regularized line--line correlators $g_M^{(\lambda)}(\chi;T;t)$ and
$g_E^{(\lambda)}(\theta;T;t)$,
the two quantities (\ref{g-lambda}), obtained {\it after} the removal of the
{\it nonperturbative} IR cutoff $T$, are still connected by the usual analytic
continuation in the angular variables only:
\ba
\overline{g}_E^{(\lambda)}(\theta;t) &=& \overline{g}_M^{(\lambda)}(i\theta;t) ,
\qquad \forall\theta\in {\cal D}_E ;
\nonumber \\
\overline{g}_M^{(\lambda)}(\chi;t) &=& \overline{g}_E^{(\lambda)}(-i\chi;t) ,
\qquad \forall\chi\in {\cal D}_M .
\label{original-ter}
\ea
For example, in {\it quenched} QED the calculation gives, in the
Minkowskian and in the Euclidean case respectively,
the following results for the fermion--fermion correlation functions
in the Feynman gauge (where the gauge--fixing parameter $\alpha$
is put equal to $1$\footnote{The free photon propagator in the generic
$\alpha$--gauge, also including the IR cutoff $\lambda$ in the form of a
photon mass, is given by:
\be
\nonumber
\widetilde{P}_{\mu\nu}(k) = -i \left( g_{\mu\nu} - (1-\alpha)
{k_\mu k_\nu \over k^2 - \alpha \lambda^2 + i\varepsilon} \right)
{1 \over k^2 - \lambda^2 + i\varepsilon} ~.
\ee}), whenever $\vec{q}_\perp \ne \vec{0}_\perp$
(i.e., $t = -|\vec{q}_\perp|^2 < 0$) \cite{Meggiolaro97}:
\ba
g_M^{f\mbox{}f}(\chi;t)^{(\lambda)} &=&
\int d^2\vec{z}_{\perp} e^{i\vec{q}_{\perp}\cdot\vec{z}_{\perp}} 
\exp\left[-ie^2|\coth\chi| \int \frac{d^2\vec{k}_{\perp}}{(2\pi)^2}
e^{i\vec{k}_{\perp}\cdot\vec{z}_{\perp}}\frac{1}{\vec{k}_{\perp}^2
+ \lambda^2}\right] ,
\nonumber \\
g_E^{f\mbox{}f}(\theta;t)^{(\lambda)} &=&
\int d^2\vec{z}_{\perp} e^{i\vec{q}_{\perp}\cdot\vec{z}_{\perp}} 
\exp\left[-e^2 {\cos\theta \over |\sin\theta|}
\int \frac{d^2\vec{k}_{\perp}}{(2\pi)^2}
e^{i\vec{k}_{\perp}\cdot\vec{z}_{\perp}}\frac{1}{\vec{k}_{\perp}^2
+ \lambda^2}\right] .
\label{QEDff}
\ea
For obtaining the fermion--antifermion correlation function it is clearly
sufficient to exchange $e^2$ with $-e^2$ in the fermion--fermion correlator,
getting:
\ba
g_M^{f\bar{f}}(\chi;t)^{(\lambda)} &=&
\int d^2\vec{z}_{\perp} e^{i\vec{q}_{\perp}\cdot\vec{z}_{\perp}} 
\exp\left[ie^2|\coth\chi| \int \frac{d^2\vec{k}_{\perp}}{(2\pi)^2}
e^{i\vec{k}_{\perp}\cdot\vec{z}_{\perp}}\frac{1}{\vec{k}_{\perp}^2
+ \lambda^2}\right] ,
\nonumber \\
g_E^{f\bar{f}}(\theta;t)^{(\lambda)} &=&
\int d^2\vec{z}_{\perp} e^{i\vec{q}_{\perp}\cdot\vec{z}_{\perp}} 
\exp\left[e^2 {\cos\theta \over |\sin\theta|}
\int \frac{d^2\vec{k}_{\perp}}{(2\pi)^2}
e^{i\vec{k}_{\perp}\cdot\vec{z}_{\perp}}\frac{1}{\vec{k}_{\perp}^2
+ \lambda^2}\right] .
\label{QEDffbar}
\ea
These correlators, seen as functions of the complex angular variables $\chi$
(Minkowskian) and $\theta$ (Euclidean), have the same analytic structure of
the loop--loop correlators discussed above. In fact, both $g_M^{ff}$ and
$g_M^{f\bar{f}}$ can be analytically extended from the positive real axis
$\chi\in\mathbb{R}^+$ to the same domain ${\cal D}_M$ defined in Eq.
(\ref{domain}) and, similarly, both $g_E^{ff}$ and $g_E^{f\bar{f}}$ can be
analytically extended from the real segment $\theta\in(0,\pi)$
to the same domain ${\cal D}_E$ defined in Eq. (\ref{domain}).
The analytic extensions $\overline{g}_M^{ff}$, $\overline{g}_M^{f\bar{f}}$,
$\overline{g}_E^{ff}$ and $\overline{g}_E^{f\bar{f}}$ are obtained from the
expressions (\ref{QEDff}) and (\ref{QEDffbar}) by the simple substitutions
$|\coth\chi| \to \coth\chi$ and $\cos\theta/|\sin\theta| \to \cot\theta$.
The analytic--continuation relations (\ref{original-ter}) are trivially
satisfied and so is the crossing--symmetry relation (\ref{eq:croce7}) or
(\ref{eq:croce8}):
\be
g_M^{f\bar{f}}(\chi;t)^{(\lambda)} = \overline{g}_M^{f\mbox{}f}
(i\pi - \chi;t)^{(\lambda)},
\qquad \forall\chi\in\mathbb{R}^+.
\label{eq:croceQED}
\ee
Let us now address the more interesting and surely more complicated question
of the computation of the line--line correlators in QCD perturbation theory.

The perturbative calculation of the quark--antiquark Minkowskian correlator
is completely analogous to the quark--quark one: as we shall see below, it
comes out that the contribution of every Feynman graph is the same as in the
quark--quark calculation, with the only difference that the colour factor
coming from the second Wilson line has to be changed according to a simple
{\it crossing} rule. In fact, when expanding each Wilson line in the
numerator of the correlation function $g_M^{qq}$ in powers of $g$ (according
to the definition of the time--ordered exponential), one finds:
\ba
\lefteqn{
g_M^{qq}(\chi;T;t)_{i'i;j'j} } \nonumber \\
& & = {1 \over [Z_M(T)]^2}
\sum_{r,s=1}^{\infty} (-ig)^{r+s}
\frac{p_1^{\mu_1}}{m} \ldots \frac{p_1^{\mu_r}}{m}
\frac{p_2^{\nu_1}}{m} \ldots \frac{p_2^{\nu_s}}{m} \nonumber \\
& & \times (T_{a_1} \ldots T_{a_r})_{i'i}
(T_{b_1} \ldots T_{b_s})_{j'j}
\nonumber \\
& & \times \int d^2\vec{z}_\perp e^{i\vec{q}_\perp \cdot \vec{z}_\perp}
\int_{-T}^{+T} d\tau_1 \ldots \int_{-T}^{+T}d\tau_r
\int_{-T}^{+T} d\sigma_1 \ldots \int_{-T}^{+T}d\sigma_s
\nonumber \\
& & \times \theta(\tau_1-\tau_2) \ldots \theta(\tau_{r-1}-\tau_r)
\theta(\sigma_1-\sigma_2) \ldots \theta(\sigma_{s-1}-\sigma_s)
\nonumber \\
& & \times
G^{a_1 \ldots a_r b_1 \ldots b_s}_{\mu_1 \ldots \mu_r \nu_1 \ldots \nu_s}
\left( z+{p_1 \over m}\tau_1, \ldots ,z+{p_1 \over m}\tau_r,
{p_2 \over m}\sigma_1, \ldots ,{p_2 \over m}\sigma_s \right) ,
\label{gMqq-exp}
\ea
having denoted with
\be
G^{a_1 \ldots a_p}_{\mu_1 \ldots \mu_p} (x_1, \ldots ,x_p) \equiv
\langle A_{\mu_1}^{a_1}(x_1) \ldots A_{\mu_p}^{a_p}(x_p) \rangle
\ee
the (complete) $p$--point gluonic Green function.\\
The analogous expansion for the quark--antiquark correlator is, of course
[see also Eq. (\ref{W2*})]:
\ba
\lefteqn{
g_M^{q\bar{q}}(\chi;T;t)_{i'i;j'j} } \nonumber \\
& & = {1 \over [Z_M(T)]^2}
\sum_{r,s=1}^{\infty} (-ig)^{r+s}
\frac{p_1^{\mu_1}}{m} \ldots \frac{p_1^{\mu_r}}{m}
\frac{p_2^{\nu_1}}{m} \ldots \frac{p_2^{\nu_s}}{m} \nonumber \\
& & \times (T_{a_1} \ldots T_{a_r})_{i'i}
(-1)^s (T_{b_1}^* \ldots T_{b_s}^*)_{j'j}
\nonumber \\
& & \times \int d^2\vec{z}_\perp e^{i\vec{q}_\perp \cdot \vec{z}_\perp}
\int_{-T}^{+T} d\tau_1 \ldots \int_{-T}^{+T}d\tau_r
\int_{-T}^{+T} d\sigma_1 \ldots \int_{-T}^{+T}d\sigma_s
\nonumber \\
& & \times \theta(\tau_1-\tau_2) \ldots \theta(\tau_{r-1}-\tau_r)
\theta(\sigma_1-\sigma_2) \ldots \theta(\sigma_{s-1}-\sigma_s)
\nonumber \\
& & \times
G^{a_1 \ldots a_r b_1 \ldots b_s}_{\mu_1 \ldots \mu_r \nu_1 \ldots \nu_s}
\left( z+{p_1 \over m}\tau_1, \ldots ,z+{p_1 \over m}\tau_r,
{p_2 \over m}\sigma_1, \ldots ,{p_2 \over m}\sigma_s \right) .
\label{gMqqbar-exp}
\ea
From a comparison with the previous relation (\ref{gMqq-exp}), we immediately
see that the same gluonic Green function
$G^{a_1 \ldots a_r b_1 \ldots b_s}_{\mu_1 \ldots \mu_r \nu_1 \ldots \nu_s}$
(and, therefore, also every given Feynman graph originated from its
perturbative expansion, taking into account also the squared renormalization
constant $[Z_M(T)]^2$ at the denominator) comes out to be contracted with
the {\it same} colour factor
$(T_{a_1} \ldots T_{a_r})_{i'i}$ for the first Wilson line and a
{\it different} colour factor $(-1)^s (T_{b_1}^* \ldots T_{b_s}^*)_{j'j}
= (-1)^s (T_{b_s} \ldots T_{b_1})_{jj'}$ for the second Wilson line [see
also Eq. (\ref{W2*}): we have used the fact that the generators $T_a$ are
hermitian]. The change
\be
(T_{b_1} \ldots T_{b_s})_{j'j} \rightarrow
(-1)^s (T_{b_s} \ldots T_{b_1})_{jj'}
\label{crossing-rule}
\ee
for each given Feynman graph from the quark--quark to the quark--antiquark
case represents what we can call the {\it ``crossing relation''} for
Feynman graphs.

Let us see, in particular, how this works in the case of correlators
evaluated in QCD perturbation theory up to the fourth order in the
(renormalized) coupling constant.
The calculation of the quark--quark Minkowskian and Euclidean correlators up
to order $g_R^4$ in the renormalized coupling constant has been already
carried out in Ref. \cite{Meggiolaro97}, with the results (always in the
Feynman gauge $\alpha = 1$):
\ba
\label{eq:nondif}
\lefteqn{
g_M^{qq}(\chi;t)^{(\lambda)}_{i'i;j'j}|_{g_R^4} } \nonumber \\
& & = ig_R^2\frac{1}{t} |\coth\chi| \left[ 1 - g_R^2\left(F^{(2)}(t)
+ \frac{2N_cB}{(4\pi)^2} + \frac{N_c}{4\pi}tI(t) |\chi| |\coth\chi|
\right)\right] \cdot (G_1)_{i'i;j'j}
\nonumber \\
& & - \frac{1}{2}g_R^4I(t)\coth^2\chi \cdot (G_2)_{i'i;j'j} ;
\ea
\ba
\label{eq:nondifE}
\lefteqn{
g_E^{qq}(\theta;t)^{(\lambda)}_{i'i;j'j}|_{g_R^4} } \nonumber \\
& & = g_R^2\frac{1}{t} {\cos\theta \over |\sin\theta|}
\left[ 1 - g_R^2\left(F^{(2)}(t)
+ \frac{2N_cB}{(4\pi)^2} + \frac{N_c}{4\pi}tI(t) \{\theta\}
{\cos\theta \over |\sin\theta|} \right)\right] \cdot (G_1)_{i'i;j'j}
\nonumber \\
& & + \frac{1}{2}g_R^4I(t)\cot^2\theta \cdot (G_2)_{i'i;j'j} ,
\qquad {\rm with:~} \{\theta\} \equiv 2 \int_0^{|\tan{\theta \over 2}|}
{dx \over 1+x^2} ,
\ea
where:
\ba
\label{G1}
(G_1)_{i'i;j'j} &\equiv& (T_a)_{i'i}(T_a)_{j'j} , \\
\label{G2}
(G_2)_{i'i;j'j} &\equiv& (T_a T_b)_{i'i}(T_a T_b)_{j'j} , \\
I(t) &\equiv& \int \frac{d^2\vec{k}_{\perp}}{(2\pi)^2}
\frac{1}{\vec{k}_{\perp}^2 + \lambda^2}
\frac{1}{(\vec{k}_{\perp}+\vec{q}_{\perp})^2 + \lambda^2} ,
\ea
while the finite constant $B$ and the function $F^{(2)}(t)$, coming from a
first--order expansion of the renormalized gluon propagator, depend on the
particular renormalization scheme adopted.
The results (\ref{eq:nondif}) and (\ref{eq:nondifE}) are valid for every
real value of $\chi$ and $\theta$.\\
(As already stressed in Ref. \cite{Meggiolaro97}, the Minkowskian result
(\ref{eq:nondif}) reproduces, in the limit of very large rapidity gap
$\chi \simeq \log(s/m^2) \to \infty$ [and only in this case!],
the well--known result obtained when computing the high--energy quark--quark
scattering amplitude with usual perturbative techniques
\cite{BFKL,Lipatov,Cheng-Wu-book}.)

The property (\ref{propM}) for the Minkowskian correlator is trivially
satisfied and, since $\{(-\theta)\}$ $= \{\theta\}$ and $\{(2\pi-\theta)\} =
\{\theta\}$, the properties (\ref{propE1}) and (\ref{propE2}) for the
Euclidean correlator are satisfied too.
Moreover, since $\{\theta\} = \theta$ for $\theta \in (0,\pi)$, one
immediately finds that the analytic extension $\overline{g}_M^{qq}$
of the Minkowskian correlator from the positive real axis $\chi\in\mathbb{R}^+$
and the analytic extension $\overline{g}_E^{qq}$ of the Euclidean correlator
from the real segment $\theta\in(0,\pi)$ are given by:
\ba
\label{eq:overgM}
\lefteqn{
\overline{g}_M^{qq}(\chi;t)^{(\lambda)}_{i'i;j'j}|_{g_R^4} } \nonumber \\
& & = ig_R^2\frac{1}{t} \coth\chi \left[ 1 - g_R^2\left(F^{(2)}(t)
+ \frac{2N_cB}{(4\pi)^2} + \frac{N_c}{4\pi}tI(t) \chi \coth\chi
\right)\right] \cdot (G_1)_{i'i;j'j}
\nonumber \\
& & - \frac{1}{2}g_R^4I(t)\coth^2\chi \cdot (G_2)_{i'i;j'j} ,
\qquad \forall\chi\in{\cal D}_M ,
\ea
\ba
\label{eq:overgE}
\lefteqn{
\overline{g}_E^{qq}(\theta;t)^{(\lambda)}_{i'i;j'j}|_{g_R^4} } \nonumber \\
& & = g_R^2\frac{1}{t} \cot\theta \left[ 1 - g_R^2\left(F^{(2)}(t)
+ \frac{2N_cB}{(4\pi)^2} + \frac{N_c}{4\pi}tI(t) \theta \cot\theta
\right)\right] \cdot (G_1)_{i'i;j'j}
\nonumber \\
& & + \frac{1}{2}g_R^4I(t)\cot^2\theta \cdot (G_2)_{i'i;j'j} ,
\qquad \forall\theta\in{\cal D}_E ,
\ea
with the usual analyticity domains ${\cal D}_M$ and ${\cal D}_E$ defined
in Eq. (\ref{domain}).
From the explicit expressions (\ref{eq:overgM}) and (\ref{eq:overgE}) one
immediately verifies that the analytic--continuation relations
(\ref{original-ter}) are verified.

Let us now turn our attention to the quark--antiquark correlator. As we have
already said above, it is not necessary to repeat the calculation from the
beginning, but we can simply use the {\it crossing relation}
(\ref{crossing-rule}) derived above to convert the contribution of
each given Feynman graph from the quark--quark to the quark--antiquark case.
In practice, it comes out that those which do not vanish, with the
exception of the two--gluon--exchange graphs {\bf b} and {\bf c} in Fig. 5,
have (apart from a multiplicative constant) the same colour factor (\ref{G1})
of the one--gluon--exchange graph {\bf a} in Fig. 5 and, apart from the
colour--index exchange $j \leftrightarrow j'$, they simply take an extra
minus sign, exactly as graph {\bf a}:
\be
(G_1)_{i'i;j'j} = (T_a)_{i'i}(T_a)_{j'j} \rightarrow
- (T_a)_{i'i}(T_a)_{jj'} = - (G_1)_{i'i;jj'} .
\label{graph-a}
\ee
(These graphs contribute to the pieces containing $B$ and $F^{(2)}(t)$ in
Eqs. (\ref{eq:nondif}) and (\ref{eq:nondifE}).  The whole set of graphs
contributing to the order ${\cal O}(g_R^4)$ are reported in Figs. 2 and 3
of Ref. \cite{Meggiolaro97}.)\\
Instead, always by virtue of the crossing relation (\ref{crossing-rule}),
graphs {\bf b} and {\bf c} simply enter in the quark--quark and
quark--antiquark correlators with exchanged colour factor (and exchanged
colour index $j \leftrightarrow j'$):
\ba
M({\rm\bf b}) \cdot (T_a T_b)_{i'i} (T_a T_b)_{j'j}
&\rightarrow&
M({\rm\bf b}) \cdot (T_a T_b)_{i'i} (T_b T_a)_{jj'} ,
\nonumber \\
M({\rm\bf c}) \cdot (T_a T_b)_{i'i} (T_b T_a)_{j'j}
&\rightarrow&
M({\rm\bf c}) \cdot (T_a T_b)_{i'i} (T_a T_b)_{jj'} .
\ea
In Ref. \cite{Meggiolaro97} it was found that:
\ba
M({\rm\bf b}) &=& i {g_R^4 \over 2\pi} I(t) (i\pi - |\chi|) \coth^2 \chi ,
\nonumber \\
M({\rm\bf c}) &=& i {g_R^4 \over 2\pi} I(t) |\chi| \coth^2 \chi ,
\ea
so that:
$M({\rm\bf b}) + M({\rm\bf c}) = -{1 \over 2} g_R^4 I(t) \coth^2 \chi$.\\
Making use of the following relation for the colours factors [with the
definitions (\ref{G1}) and (\ref{G2})]:
\ba
(T_a T_b)_{ij} (T_b T_a)_{kl} &=&
(T_a T_b)_{ij} (T_a T_b)_{kl} + {N_c \over 2} (T_c)_{ij} (T_c)_{kl}
\nonumber \\
&\equiv& (G_2)_{ij;kl} + {N_c \over 2} (G_1)_{ij;kl} ,
\ea
we obtain that the contribution of graphs {\bf b} and {\bf c} to the
quark--antiquark correlator is given by the following expression:
\ba
\lefteqn{
M({\rm\bf b}) \cdot (T_a T_b)_{i'i} (T_b T_a)_{jj'} +
M({\rm\bf c}) \cdot (T_a T_b)_{i'i} (T_a T_b)_{jj'} } \nonumber \\
& & = {N_c \over 2} M({\rm\bf b}) \cdot (G_1)_{i'i;jj'} +
[ M({\rm\bf b}) + M({\rm\bf c}) ] \cdot (G_2)_{i'i;jj'} \nonumber \\
& & = i {N_c g_R^4 \over 4\pi} I(t) (i\pi - |\chi|) \coth^2 \chi
\cdot (G_1)_{i'i;jj'}
-{1 \over 2} g_R^4 I(t) \coth^2 \chi \cdot (G_2)_{i'i;jj'} .
\ea
Summing all contributions, the following result is found for the
quark--antiquark correlator at order ${\cal O}(g_R^4)$, for positive
hyperbolic angle $\chi > 0$:
\ba
\label{eq:nondif2}
\lefteqn{
g_M^{q\bar{q}}(\chi;t)^{(\lambda)}_{i'i;j'j}|_{g_R^4} } \nonumber \\
& & = -ig_R^2\frac{1}{t}\coth\chi \left[1 - g_R^2\left(F^{(2)}(t)
+\frac{2N_cB}{(4\pi)^2} - \frac{N_c}{4\pi}tI(t)(i\pi-\chi)\coth\chi
\right)\right]\cdot (G_1)_{i'i;jj'}
\nonumber \\
& & - \frac{1}{2}g_R^4I(t)\coth^2\chi \cdot (G_2)_{i'i;jj'},
\qquad \forall\chi\in\mathbb{R}^+.
\ea
This is also the expression of the analytic extension
$\overline{g}_M^{q\bar{q}}$ from the real positive $\chi$--axis to the
{\it same} analyticity domain
${\cal D}_M = \{\chi \in \mathbb{C} | \chi \ne ik\pi,~ k \in \mathbb{Z}\}$
introduced above for the quark--quark correlator $\overline{g}_M^{qq}$
written in Eq. (\ref{eq:overgM}). A comparison of (\ref{eq:nondif2}) with
(\ref{eq:overgM}) shows that the crossing--symmetry relation is verified:
\be
g_M^{q\bar{q}}(\chi;t)^{(\lambda)}_{i'i;j'j}|_{g_R^4} =
\overline{g}_M^{qq}(i\pi-\chi;t)^{(\lambda)}_{i'i;jj'}|_{g_R^4} ,
\qquad \forall\chi\in\mathbb{R}^+.
\ee
Going to larger perturbative orders, many more and much more complicated
Feynman diagrams are involved: however, there is apparently no reason
why the results that we have found and discussed above concerning
analyticity and crossing symmetry should not be true also in these cases.
In the appendix A, for example, we discuss correlators at orders
${\cal O}(g_R^6)$, limiting ourselves (for simplicity) to the (physically
interesting) {\it diffractive} part (defined according to Ref.
\cite{Nachtmann91}) of the three--gluon--exchange graphs
(diagrams {\bf d} $\div$ {\bf i} in Fig. 5).

\newpage

\newsection{Concluding remarks and prospects}

\noindent
The main result of this paper has been to clarify the relation between
Minkowskian--to--Euclidean analyticity properties and crossing symmetry
both for line--line and loop--loop eikonal amplitudes.
In sections 2 and 3 we have shown, in a nonperturbative way, using the
functional integral approach, how certain apparently {\it reasonable}
analyticity hypotheses of the line--line and loop--loop correlation
functions in the angular variables and in the IR cutoff $T$, which are known
to imply the Minkowskian--to--Euclidean analytic--continuation relation, also
imply (directly from this) the crossing--symmetry relation, of which a nice
geometrical interpretation has been provided.
The {\it reasonableness} of the above--mentioned analyticity hypotheses
comes essentially from the explicit tests that have been done in
perturbation theory and which are presented in section 4 and appendix A.
Of course, as we have already said at the beginning of section 4,
perturbation theory is the only available technique for computing (from
first principles) {\it both} the Minkowskian {\it and} the Euclidean
correlation functions and so explicitly testing all the above--mentioned
analyticity and crossing--symmetry properties.

A real nonperturbative foundation of these properties is at the moment
out of our reach, but this is really the kind of effort that one should
make in order to fully understand (and so fully trust!) the nonperturbative
results which derive from them.
(In appendix B, for example, we have discussed the connection between
some singularities in the correlators and the emergence of certain
parton--parton or dipole--dipole bound states. It would be nice if one
could generalize this kind of arguments and find a nonperturbative
way of identifying all type of singularities in the correlators and so
have a complete description of their analyticity structure.)

There exist in the literature some nonperturbative computations of the
Euclidean correlation functions, obtained using some specific models
in the Euclidean theory. (They can then be continued to the corresponding
Minkowskian correlation functions using the analytic--continuation
relation in the angular variables and in the IR cutoff $T$ and so one
can in principle address, from a fully nonperturbative point of view,
the formidable [and unfortunately still unsolved!] problem of the
asymptotic $s$--dependence of hadron--hadron scattering amplitudes and
total cross sections.)

For example, in Ref. \cite{LLCM2} the loop--loop Euclidean correlation
functions have been evaluated in the context of the so--called
{\it ``loop--loop correlation model''} \cite{LLCM1}, in which the QCD
vacuum is described by perturbative gluon exchange and the nonperturbative
{\it ``Stochastic Vacuum Model''}: the result is a loop--loop Euclidean
correlation function ${\cal C}_E (\theta;\vec{z}_\perp,\vec{R}_{1\perp},
\vec{R}_{2\perp})$ which, for $\theta \in (0,\pi)$, is an analytic
function of $\cot \theta$ and can then be analytically extended to the
entire complex plane with the exception of the singularities of $\cot\theta$,
i.e., to the same domain given in Eq. (\ref{domain}),
${\cal D}_E = \{\theta \in \mathbb{C}| \theta \ne k\pi,~ k\in \mathbb{Z}\}$
[including the real segment $(0,\pi)$, the negative imaginary axis
$\theta = -i\chi$, $\chi > 0$, and the semiaxis ($\re\theta=\pi,\im\theta>0$)].
The Euclidean--to--Minkowskian analytic continuation can
then be safely applied and the crossing--symmetry relation comes out to be
trivially satisfied.

The same also happens adopting a different Euclidean approach
\cite{instanton1}, consisting in evaluating the {\it one--instanton}
contribution to both the line--line (see also Ref. \cite{instanton2}) and the
loop--loop Euclidean correlation functions: one finds that the colour--elastic
line--line and loop--loop Euclidean correlation functions for $\theta
\in (0,\pi)$ scale as $1/\sin\theta$, while the colour--changing inelastic
line--line Euclidean correlation function scales as $\cot\theta$.

Finally, we want to comment on a third Euclidean approach existing in the
literature, in which one computes the line--line/loop--loop Euclidean
correlation functions in strongly coupled gauge theories using the
AdS/CFT correspondence \cite{JP1,JP2,Janik}.
In a first paper \cite{JP1} this approach was used to study the loop--loop
Euclidean correlation function in the ${\cal N}=4$ SYM theory in the
limit of large number of colours ($N_c \to \infty$) and strong coupling:
one finds that the Euclidean correlation function for $\theta\in (0,\pi)$
is a combination of various pieces scaling as $1/\sin\theta$, $\cot\theta$
and $\cos^2\theta/\sin\theta$, which can be analytically extended to the
same complex domain ${\cal D}_E$ which was considered in the two previous
cases and in Eq. (\ref{domain}).

A different situation appears, instead, when one tries to extend the
approach based on the AdS/CFT correspondence in order to study the
line--line/loop--loop Euclidean correlation functions in strongly
coupled {\it confining} (i.e., nonconformal) gauge theories, as was done
in Refs. \cite{JP2,Janik}. In this case the analytic structure of the
Euclidean correlation functions involves branch cuts in the complex $\theta$
and $T$ planes, coming from logarithms and square roots, which lead to
an ambiguity in the Euclidean--to--Minkowskian analytic continuation.
It is not clear to us, at the moment, if this analytic structure is
something peculiar to the specific model considered there or, vice versa,
if it is a more general characteristic, maybe related to the presence
of confinement (as it seems to be indicated by the authors in
Refs. \cite{JP2,Janik}).
In the case of this second possibility, one is immediately faced with the
following series of questions:
\begin{itemize}
\item[i)] Are the analyticity hypotheses considered in section 2 maybe too
strong and to what degree can they be relaxed?
\item[ii)] What about the Euclidean--to--Minkowskian analytic continuation
if weaker analyticity hypotheses are kept in place of those discussed in
section 2?
\item[iii)] What about, finally, the crossing symmetry relation in the
presence of a more complicated analytic structure, when the
Euclidean--to--Minkowskian analytic continuation cannot be trusted, at least
in the form presented in section 2?
\end{itemize}
We have not, at the moment, the answers for all these questions but we hope
that future work along the lines indicated in this paper will shed some
light on these and other related problems.


\newpage

\renewcommand{\thesection}{}
\renewcommand{\thesubsection}{A.\arabic{subsection}}

\pagebreak[3]
\setcounter{section}{1}
\setcounter{equation}{0}
\setcounter{subsection}{0}
\setcounter{footnote}{0}

\begin{flushleft}
{\Large\bf \thesection Appendix A: Three--gluon--exchange diffractive
contributions to $qq$ and $q\bar{q}$ correlators}
\end{flushleft}

\renewcommand{\thesection}{A}

\noindent
Order ${\cal O}(g_R^6)$ calculations involve much more, and more
complicated, Feynman graphs than order ${\cal O}(g_R^4)$.
In this appendix we shall limit ourselves to the study of the
{\it diffractive} part of the three--gluon--exchange graphs only.
The {\it diffractive} parts of the quark--quark and quark--antiquark
correlators are defined according to Ref. \cite{Nachtmann91} as:
\ba
g_M^{qq}(\chi;t)^{(D)} &\equiv&
\sum_{i,j = 1}^{N_c}g_M^{qq}(\chi;t)_{ii;jj} ,
\nonumber \\  
g_M^{q\bar{q}}(\chi;t)^{(D)} &\equiv&
\sum_{i,j = 1}^{N_c}g_M^{q\bar{q}}(\chi;t)_{ii;jj} ,
\ea
i.e., as the traces over the colour factors of each line,
and thus correspond to a process without exchange of colour.
(By virtue of Eq. (\ref{scatt}) and of the optical theorem, their real parts,
in the limit of very large rapidity gap $\chi\simeq\log(s/m^2)\to\infty$ and
vanishing squared transferred momentum $t\to 0$, are related to the
colour--averaged quark--quark and quark--antiquark total cross sections
at high energies.)

Because of the cyclicity property of the trace, diagrams differing only by an
even permutation of vertices have the same colour factor.
Therefore, the diffractive contributions of the six three--gluon--exchange
diagrams {\bf d} $\div$ {\bf i} in Fig. 5 have only two different colour
factors, one ($S_3$) for the {\it ``ladder''} diagrams {\bf d}, {\bf e},
{\bf f}, and another ($S'_3$) for the {\it ``crossed''} diagrams {\bf g},
{\bf h}, {\bf i}, given by:
\ba
S_3 &\equiv& {\rm Tr}\left[T_a T_b T_c\right]{\rm Tr}\left[T_a T_b T_c\right]
=-\frac{N_c^2 -1}{4N_c} ,\\
S_3' &\equiv& {\rm Tr}\left[T_a T_b T_c\right]{\rm Tr}\left[T_b T_a T_c\right]
= S_3 + \frac{N_c}{2}S_2 ,
\ea
where:
\be
S_2 \equiv
\textrm{Tr}\left[T_a T_b\right]\textrm{Tr}\left[T_a T_b\right]
= \frac{N_c^2 -1}{4} .
\ee
The diffractive contributions of the six diagrams to the quark--quark
correlator $g^{qq}_M$ can then be written in the form:
\ba
\Delta g^{qq}_M(\chi,t)^{(D)}_{3{\rm gluon}} &=& S_3 L(\chi,t) + S_3' X(\chi,t)
\nonumber \\
&=& S_3\left[L(\chi,t)+X(\chi,t)\right] + \frac{N_c}{2}S_2 X(\chi,t).
\ea
The diffractive contribution of the same six diagrams to the quark--antiquark
correlator $g^{q\bar{q}}_M$ is, according to the crossing relation
(\ref{crossing-rule}):
\ba
\Delta g^{q\bar{q}}_M(\chi,t)^{(D)}_{3{\rm gluon}} &=&
(-1)^3\left[S_3' L(\chi,t) + S_3 X(\chi,t)\right] \nonumber \\
&=& -S_3\left[L(\chi,t)+X(\chi,t)\right] - \frac{N_c}{2}S_2 L(\chi,t).
\ea
The three--gluon--exchange contributions are given by the expressions
(for $\chi\in\mathbb{R}^+$):
\ba
\label{qq-3g}
\lefteqn{\Delta g_M^{qq}(\chi;t)^{(D)}_{3{\rm gluon}} = } \nonumber \\
& & ig_R^6 \coth^3\chi \left\{S_3\left[\frac{1}{6}I_1(t)\right] +
\frac{N_c}{2}S_2\left[\frac{i}{2\pi} \left(\chi - i\frac{2\pi}{3}\right)I_1(t)
+ \frac{1}{2\pi^2}H(\chi)\right]\right\} ,\\
\label{qqbar-3g}
\lefteqn{\Delta g_M^{q\bar{q}}(\chi;t)^{(D)}_{3{\rm gluon}} = } \nonumber \\
& & -ig_R^6 \coth^3\chi \left\{S_3\left[\frac{1}{6}I_1(t)\right] -
\frac{N_c}{2}S_2\left[\frac{i}{2\pi}\left(\chi - i\frac{\pi}{3}\right)I_1(t)
+ \frac{1}{2\pi^2}H(\chi)\right]\right\} ,
\ea
where:
\be
I_1(t) \equiv \int \frac{d^2\vec{k}_{1\perp}}{(2\pi)^2}
\frac{d^2\vec{k}_{2\perp}}{(2\pi)^2}
\frac{1}{(\vec{k}_{1\perp})^2 + \lambda^2}
\frac{1}{(\vec{k}_{2\perp})^2 + \lambda^2}
\frac{1}
{(\vec{q}_{\perp} - \vec{k}_{1\perp} - \vec{k}_{2\perp})^2 + \lambda^2} ,
\ee
and the function $H(\chi)$ is defined by the integral:
\ba
H(\chi)&=& \int\frac{d^2\vec{k}_{1\perp}}{(2\pi)^2}
\frac{d^2\vec{k}_{2\perp}}{(2\pi)^2}
\frac{d^2\vec{k}_{3\perp}}{(2\pi)^2}(2\pi)^2
\delta^{(2)} \left(\vec{q}_{\perp} -\vec{k}_{1\perp}
-\vec{k}_{2\perp} - \vec{k}_{3\perp}\right)
\nonumber \\
&\times& h\left(\chi;\vec{k}_{1\perp},\vec{k}_{2\perp},
\vec{k}_{3\perp}\right) ,\nonumber \\
h\left(\chi;\vec{k}_{1\perp},\vec{k}_{2\perp},\vec{k}_{3\perp}\right)
&=& \int_{-\infty}^{+\infty}d\xi \int_{-\infty}^{+\infty} d\eta\,
P\frac{1}{\xi} P\frac{1}{\eta} \nonumber \\
&\times& \left(\frac{1}{\xi^2 + \vec{k}_{1\perp}^{\,2} + \lambda^2}
\frac{1}{\eta^2 + \vec{k}_{2\perp}^{\,2} + \lambda^2} -
\frac{1}{\vec{k}_{1\perp}^{\,2}+\lambda^2}
\frac{1}{\vec{k}_{2\perp}^{\,2}+\lambda^2}\right) \nonumber \\
&\times& \frac{1}{\xi^2 + \eta^2 - 2\xi\eta\cosh\chi+ \vec{k}_{3\perp}^{\,2}
+ \lambda^2  - i\epsilon} ,
\label{eq:enorme}
\ea
having denoted with
\be
P\frac{1}{\xi} \equiv \frac{1}{2} 
\left( \frac{1}{\xi -i\epsilon}+ \frac{1}{\xi + i\epsilon} \right)
\ee
the {\it ``Cauchy principal part''} of $1/\xi$.
The explicit form of $H(\chi)$ is not too enlightening; anyway, it can be
continued analytically from the positive real axis $\chi\in\mathbb{R}^+$ into
a domain including also the imaginary segment $({\re}\chi=0, 0<{\im}\chi< \pi)$
and the semiaxis $({\re}\chi<0, {\im}\chi = \pi)$. Using the notation
previously introduced, we denote such an extension as $\overline{H}(\chi)$;
as one immediately sees from (\ref{eq:enorme}), it has the property:
\be
\label{eq:simme}
\overline{H}(i\pi - \chi) = -\overline{H}(\chi).
\ee
Repeating the calculation in the Euclidean case, we obtain the result,
for $\theta\in(0,\pi)$:
\ba
\lefteqn{\Delta g_E^{qq}(\theta;t)^{(D)}_{3{\rm gluon}} = } \nonumber \\
& & -g_R^6 \cot^3\theta\left\{S_3\left[\frac{1}{6}I_1(t)\right] +
\frac{N_c}{2}S_2\left[\frac{i}{2\pi} \left( i\theta - i\frac{2\pi}{3} \right)
I_1(t) +\frac{1}{2\pi^2}\overline{H}(i\theta)\right]\right\} ,
\ea
immediately verifying the analytic--continuation relation:
\be
\Delta g_E^{qq}(\theta;t)^{(D)}_{3{\rm gluon}} =
\Delta \overline{g}_M^{qq}(i\theta;t)^{(D)}_{3{\rm gluon}}.
\label{3g-analytic}
\ee
Moreover, by virtue of the property (\ref{eq:simme}), the contributions
(\ref{qq-3g}) and (\ref{qqbar-3g}) of the three--gluon--exchange diagrams to
the $qq$ and $q\bar{q}$ correlators satisfy the crossing--symmetry relation:
\be
\Delta g_M^{q\bar{q}}(\chi;t)^{(D)}_{3{\rm gluon}}
= \Delta \overline{g}_M^{qq}(i\pi-\chi;t)^{(D)}_{3{\rm gluon}} .
\label{3g-crossing}
\ee
The three--gluon--exchange diffractive contributions to the $qq$ and $q\bar{q}$
eikonal scattering amplitudes are readily obtained once we
know the asymptotic behaviour of the function $H(\chi)$ in the limit
of very large rapidity gap $\chi\simeq\log(s/m^2)\to\infty$.
From the explicit expressions (\ref{eq:enorme}) reported above, it is easy
to see that the derivative $dH(\chi)/d\chi$ tends to zero when $\chi$ goes
to infinity and, therefore, $H(\chi)$ tends to a constant in the same limit:
\be
\lim_{\chi\to +\infty} H(\chi) = H_0.
\ee
The high--$\chi$ diffractive contribution of the three--gluon--exchange
diagrams to the quark--quark correlator is then:
\ba
\lefteqn{
\Delta g_M^{qq}(\chi\simeq\log(s/m^2)\to+\infty;t)^{(D)}_{3{\rm gluon}} }
\nonumber \\
& & \simeq ig_R^6 \left\{S_3 \left[\frac{1}{6}I_1(t) \right]
+ \frac{N_c}{2}S_2\left[\frac{i}{2\pi}\left(\log\left({s \over m^2}\right)
-i\frac{2\pi}{3}\right) I_1(t) + \frac{1}{2\pi^2}H_0\right]\right\}
\nonumber \\
& & = -g_R^6\frac{N_c S_2}{4\pi} I_1(t) \log\left({s \over m^2}\right)
 + {\rm constant~imaginary~part},
\label{qq-3g-asympt}
\ea
while the high--$\chi$ diffractive contribution of the same diagrams to the
quark--antiquark correlator is:
\ba
\lefteqn{
\Delta g_M^{q\bar{q}}(\chi\simeq\log(s/m^2)\to+\infty;t)^{(D)}_{3{\rm gluon}} }
\nonumber \\
& & \simeq -ig_R^6\left\{S_3\left[\frac{1}{6}I_1(t)\right] -
\frac{N_c}{2}S_2\left[\frac{i}{2\pi}\left(\log\left({s \over m^2}\right)
-i\frac{\pi}{3}\right) I_1(t) + \frac{1}{2\pi^2}H_0\right]\right\}
\nonumber \\
& & = -g_R^6\frac{N_c S_2}{4\pi} I_1(t) \log\left({s \over m^2}\right)
 + {\rm constant~imaginary~part}.
\label{qqbar-3g-asympt}
\ea
As in the case of the (full) ${\cal O}(g_R^4)$ results discussed in section 4,
also the (partial) ${\cal O}(g_R^6)$ results (\ref{qq-3g-asympt}) and
(\ref{qqbar-3g-asympt}) agree with the corresponding (i.e.,
three--gluon--exchange and diffractive) results obtained when computing the
high--energy quark--quark and quark--antiquark scattering amplitudes with
usual perturbative techniques \cite{BFKL,Lipatov,Cheng-Wu-book}.
This agreement as well as the analytic--continuation relation
(\ref{3g-analytic}) and the crossing--symmetry relation (\ref{3g-crossing})
are, of course, expected to hold also for all other ${\cal O}(g_R^6)$
contributions, i.e., for the full ${\cal O}(g_R^6)$ results.
As we have already said in section 4, in the case of the loop--loop
correlation function the full ${\cal O}(g_R^6)$ perturbative expansion
has been computed in Ref. \cite{BB} and found to be in agreement with both
the BFKL results \cite{BFKL,Lipatov,Cheng-Wu-book} (in the limit ov very
large rapidity gap) and with the analytic--continuation relation (\ref{final}).

\newpage

\renewcommand{\thesection}{}
\renewcommand{\thesubsection}{B.\arabic{subsection}}

\pagebreak[3]
\setcounter{section}{1}
\setcounter{equation}{0}
\setcounter{subsection}{0}
\setcounter{footnote}{0}

\begin{flushleft}
{\Large\bf \thesection Appendix B: Angular singularities {\it vs.} bound
states}
\end{flushleft}

\renewcommand{\thesection}{B}

\noindent
In this appendix we want to show that the singularities of the Euclidean
correlation functions (when $T \to \infty$) in the points $\theta = k\pi$,
$k\in\mathbb{Z}$, that we have found in all examples decribed in section 4,
are indeed expected on general (i.e., nonperturbative) grounds as the
consequence of the relation of these quantities with the potential of
certain static dipole--dipole or parton--parton bound states.

For example, considering in particular the loop--loop Euclidean correlation
function ${\cal G}_E(\theta;T;\vec{z}_\perp,\vec{R}_{1\perp},
\vec{R}_{2\perp})$, it is well known that in the case $\theta = 0$ and
$T \to \infty$ this quantity is related to the {\it van der Waals} potential
$V_{12}(\vec{z}_\perp,\vec{R}_{1\perp},\vec{R}_{2\perp})$ between two static
fermion--antifermion dipoles (one positioned in $\vec{z}_\perp \pm
\vec{R}_{1\perp}/2$ and the other positioned in $\pm \vec{R}_{2\perp}/2$)
by the following expression \cite{dipole1,dipole2}:
\be
{\cal G}_E(\theta = 0;T;\vec{z}_\perp,\vec{R}_{1\perp},\vec{R}_{2\perp})
\mathop\simeq_{T \to \infty} \exp \left[
-2T~ V_{12}(\vec{z}_\perp,\vec{R}_{1\perp},\vec{R}_{2\perp}) \right] .
\label{GE-0}
\ee
This van der Waals potential can be studied both in QCD perturbation theory
\cite{dipole1,dipole2,dipole3} and with nonperturbative techniques (see, e.g.,
Ref. \cite{LLCM2} and references therein).\\
As a pedagogic example, in {\it quenched} QED this quantity can be easily
calculated from the expressions for
${\cal G}_E = \exp[-(\widetilde{\Phi}_1^{(T)} + \widetilde{\Phi}_2^{(T)})]$
reported in Ref. \cite{Meggiolaro05}, Eqs. (2.10),
(2.11) and (2.13), where we have to put $\theta = 0$.
It is immediate to see that the integral $\widetilde{\Phi}_2^{(T)}$ (with
$\theta = 0$) continues to vanish in the large--$T$ limit, while the integral
$\widetilde{\Phi}_1^{(T)}$ behaves, for $\theta = 0$ and in the large--$T$
limit, exactly as $2T~V_{12}$ where:
\ba
V_{12}(\vec{z}_\perp,\vec{R}_{1\perp},\vec{R}_{2\perp}) &=&
{e^2 \over 4\pi} \left( 
{1 \over \vert \vec{z}_\perp+{\vec{R}_{1\perp} \over 2}
-{\vec{R}_{2\perp} \over 2} \vert}
+{1 \over \vert \vec{z}_\perp-{\vec{R}_{1\perp} \over 2}
+{\vec{R}_{2\perp} \over 2} \vert} \right.
\nonumber \\
&-& \left. {1 \over \vert \vec{z}_\perp+{\vec{R}_{1\perp} \over 2}
+{\vec{R}_{2\perp} \over 2} \vert}
-{1 \over \vert \vec{z}_\perp-{\vec{R}_{1\perp} \over 2}
-{\vec{R}_{2\perp} \over 2} \vert} \right) .
\label{potential-loop}
\ea
This is indeed the electromagnetic van der Waals potential between two static
fermion--antifermion dipoles, in the {\it quenched} approximation.

Coming back to the more general case, Eq. (\ref{GE-0}) tells us that the
correlator ${\cal G}_E$ when $T \to \infty$ has a singularity in $\theta = 0$.
The use of the crossing--symmetry relation (\ref{eq:croce15}) then immediately
tells us that ${\cal G}_E$ when $T \to \infty$ has also a singularity in
$\theta = \pi$ and therefore, by virtue of the periodicity in $\theta$,
a singularity is expected in each point $\theta = k\pi$, $k\in\mathbb{Z}$.\\
A similar result (obtained using a similar approach) is expected to hold
also for the quark--quark and quark--antiquark Euclidean correlation functions.
(The Euclidean colour--singlet quark--antiquark correlator at $\theta = 0$
with impact parameter $\vec{z}_\perp$ is essentially the expectation value of
a single Euclidean Wilson loop with transverse separation $\vec{z}_\perp$.
This quantity is known to be related, in the large--$T$ limit, to the potential
$V_{q\bar{q}}(\vec{z}_\perp)$ between a static quark and a static
antiquark separated by $\vec{z}_\perp$ \cite{Wilson74,BW79}.)

\newpage



\newpage

\noindent
\begin{center}
{\large\bf FIGURE CAPTIONS}
\end{center}
\vskip 0.5 cm
\begin{itemize}
\item [\bf Fig.~1.] The space--time configuration of the two Wilson
lines $W_{p_1}$ and $W_{p_2}$ entering in the expression for
the quark--quark elastic scattering amplitude in the high--energy limit.
\item [\bf Fig.~2.] The sequence of Euclidean transformations (leaving the
functional integral unchanged) which shows how Eq. (\ref{eq:croce4}) can
be written in the form (\ref{eq:croce5}).
\item [\bf Fig.~3.] The common analyticity domain of $\overline{g}_E^{qq}$
and $\overline{g}_E^{q\bar{q}}$ in the complex variable $\theta$.
\item [\bf Fig.~4.] The common analyticity domain of $\overline{g}_M^{qq}$
and $\overline{g}_M^{q\bar{q}}$ in the complex variable $\chi$.
\item [\bf Fig.~5.] The Feynman diagrams with exchange of one, two and three
gluons which contribute to the quark--quark and quark--antiquark correlators.
\end{itemize}

\newpage

\pagestyle{empty}

\centerline{\large\bf Figure 1}
\vskip 4truecm
\begin{figure}[htb]
\vskip 4.5truecm
\includegraphics{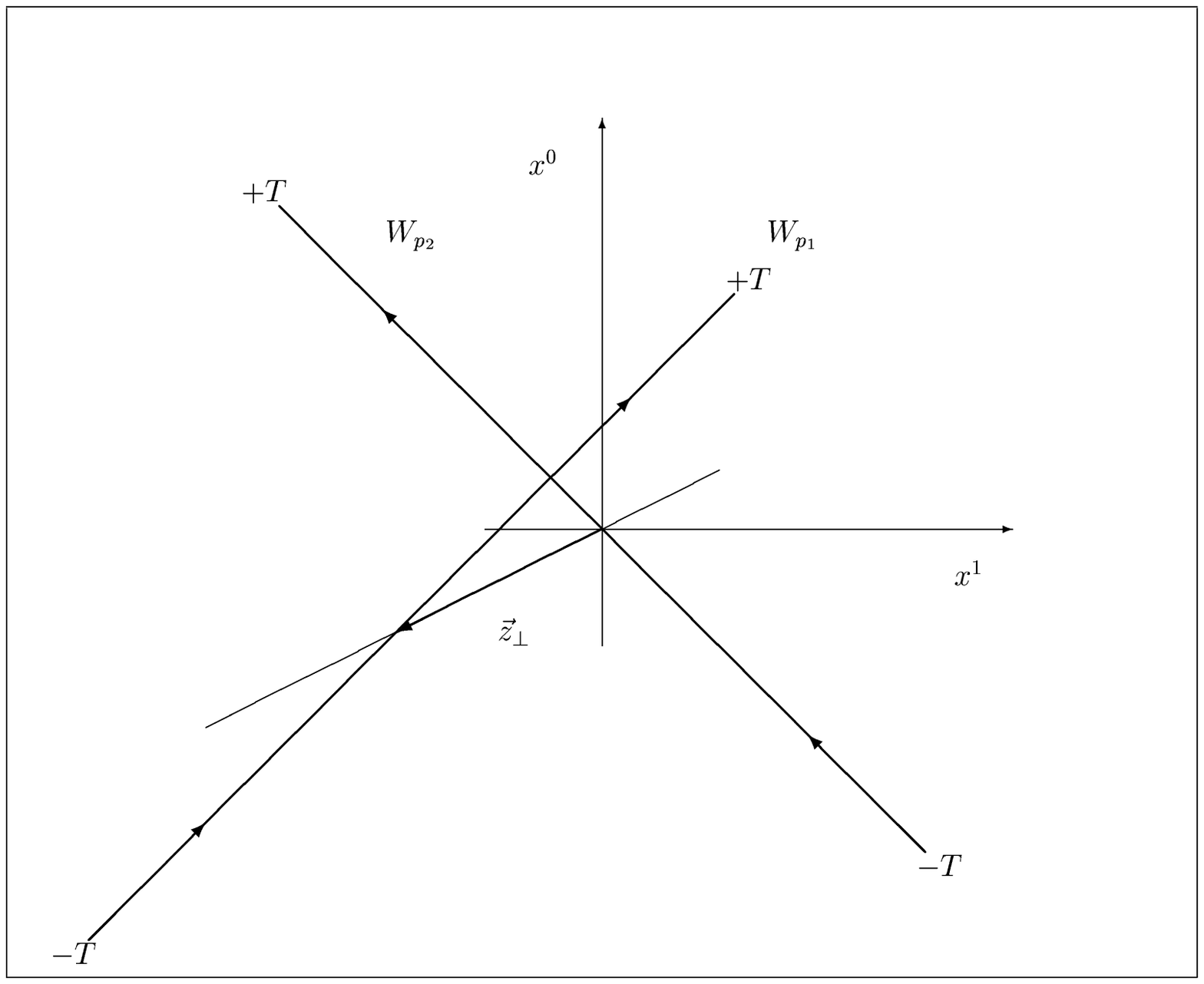}
\label{fig1}
\end{figure}
\vskip 9.5 cm
\begin{itemize}
\item [\bf Fig.~1.] The space--time configuration of the two Wilson
lines $W_{p_1}$ and $W_{p_2}$ entering in the expression for
the quark--quark elastic scattering amplitude in the high--energy limit.
\end{itemize}

\newpage

\centerline{\large\bf Figure 2}
\vskip 4truecm
\begin{figure}[htb]
\vskip 4.5truecm
\includegraphics{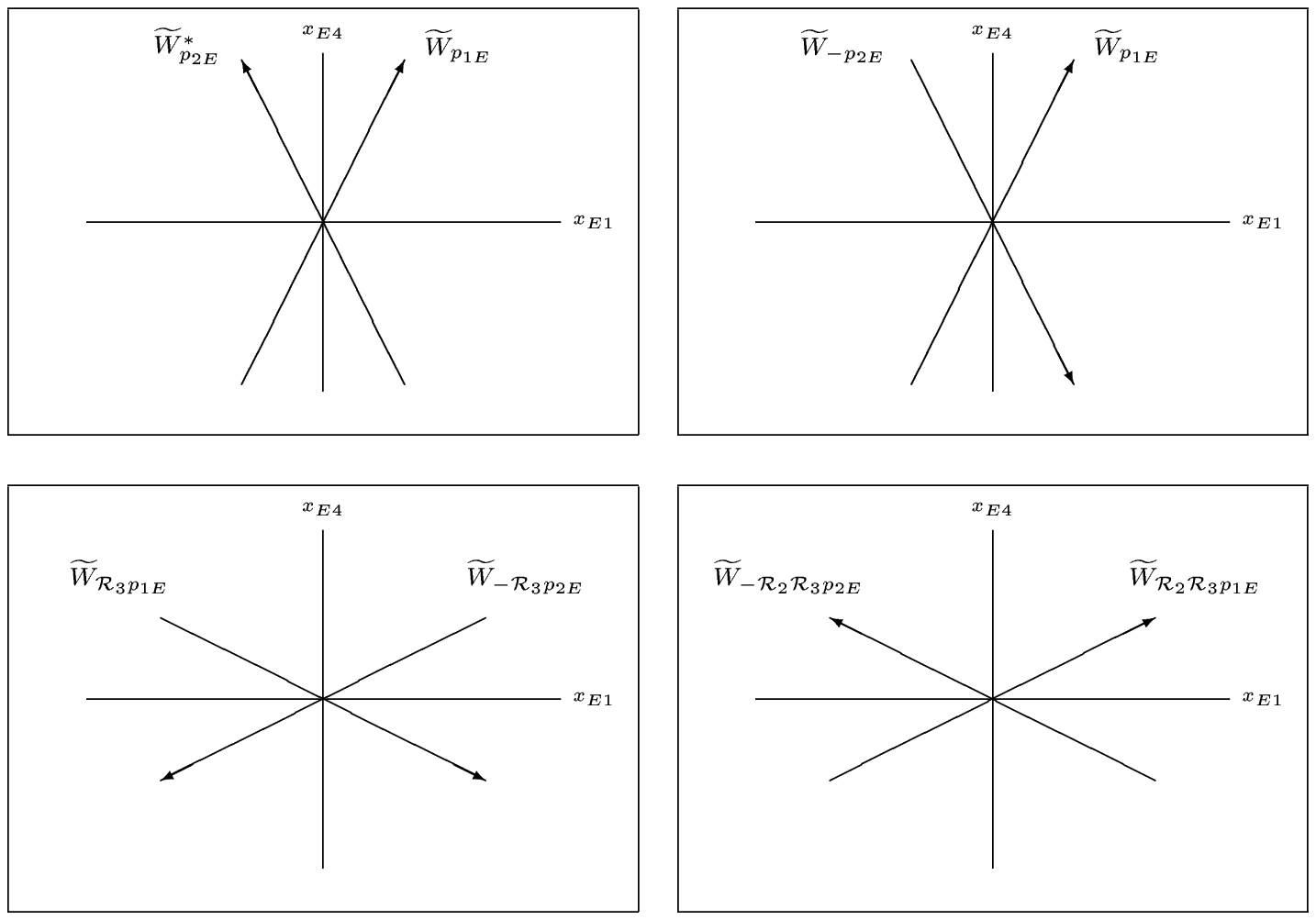}
\label{fig2}
\end{figure}
\vskip 9.5 cm
\begin{itemize}
\item [\bf Fig.~2.] The sequence of Euclidean transformations (leaving the
functional integral unchanged) which shows how Eq. (\ref{eq:croce4}) can
be written in the form (\ref{eq:croce5}).
\end{itemize}

\newpage

\centerline{\large\bf Figure 3}
\vskip 4truecm
\begin{figure}[ht]
\centering
\includegraphics[height= 0.7\textwidth, width=1\textwidth]{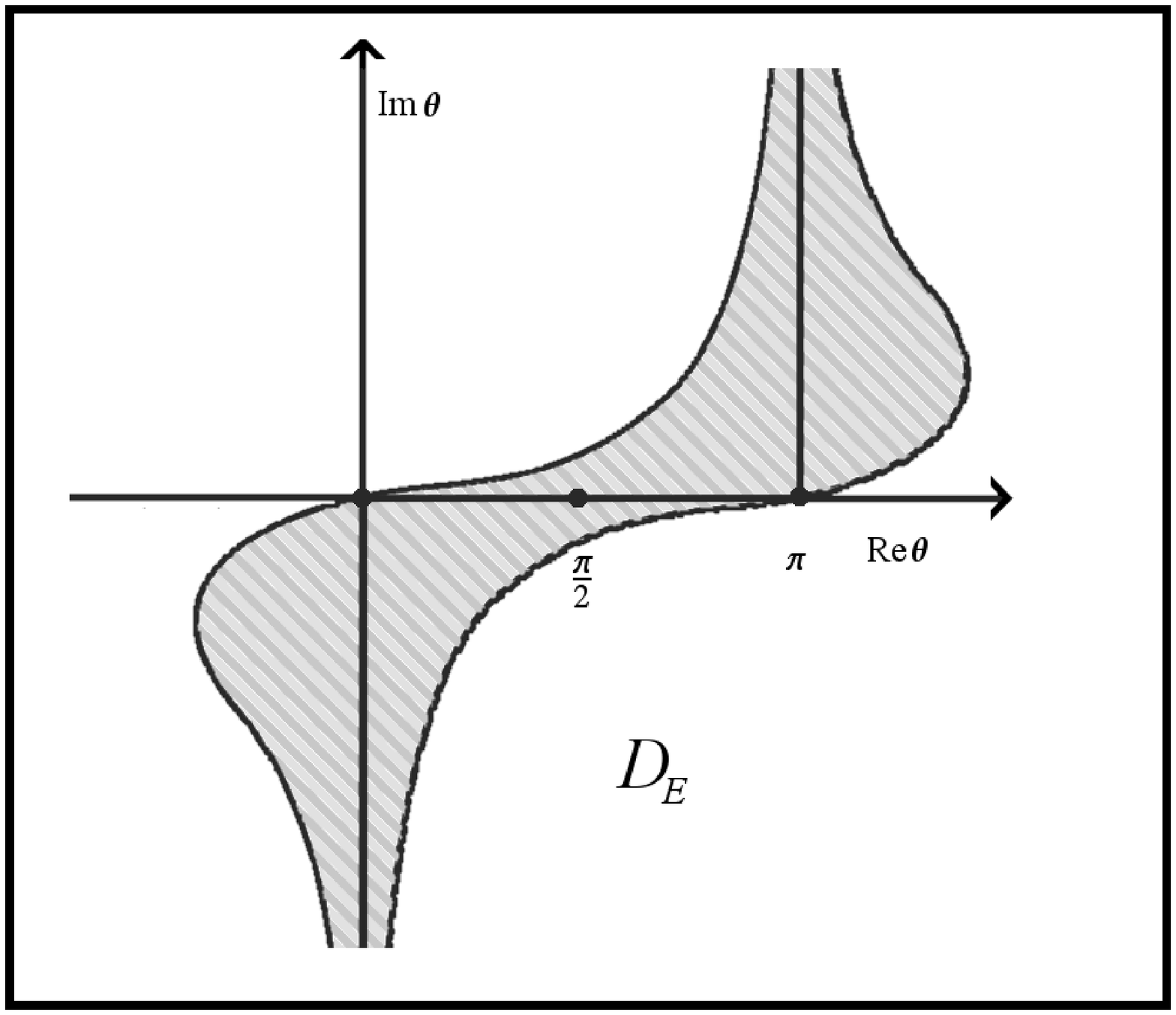}
\label{fig3}
\end{figure}
\begin{itemize}
\item [\bf Fig.~3.] The common analyticity domain of $\overline{g}_E^{qq}$
and $\overline{g}_E^{q\bar{q}}$ in the complex variable $\theta$.
\end{itemize}

\newpage

\centerline{\large\bf Figure 4}
\vskip 4truecm
\begin{figure}[ht]
\centering
\includegraphics[height= 0.7\textwidth, width=1\textwidth]{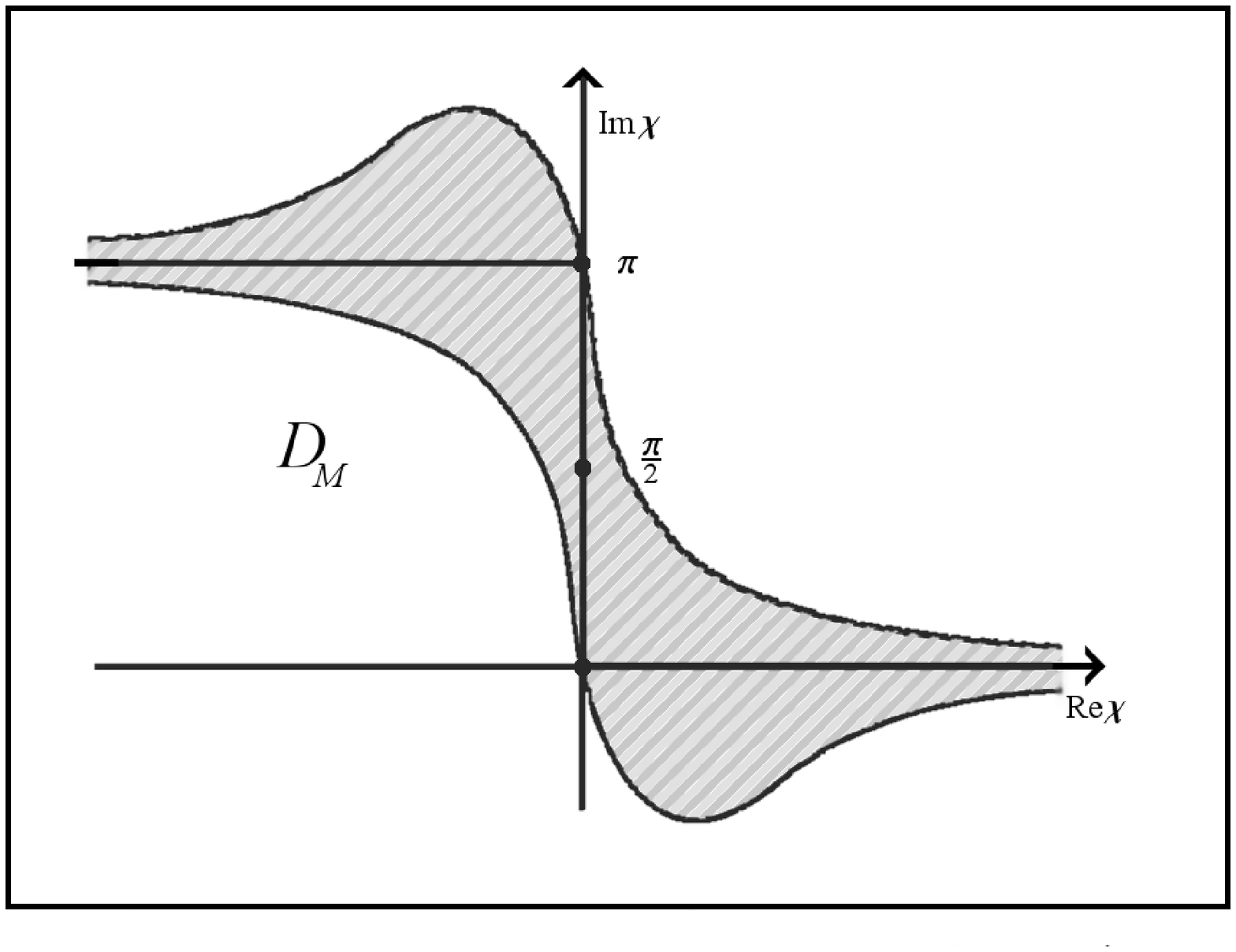}
\label{fig4}
\end{figure}
\begin{itemize}
\item [\bf Fig.~4.] The common analyticity domain of $\overline{g}_M^{qq}$
and $\overline{g}_M^{q\bar{q}}$ in the complex variable $\chi$.
\end{itemize}

\newpage

\centerline{\large\bf Figure 5}
\vskip 4truecm
\begin{figure}[htb]
\vskip 4.5truecm
\includegraphics{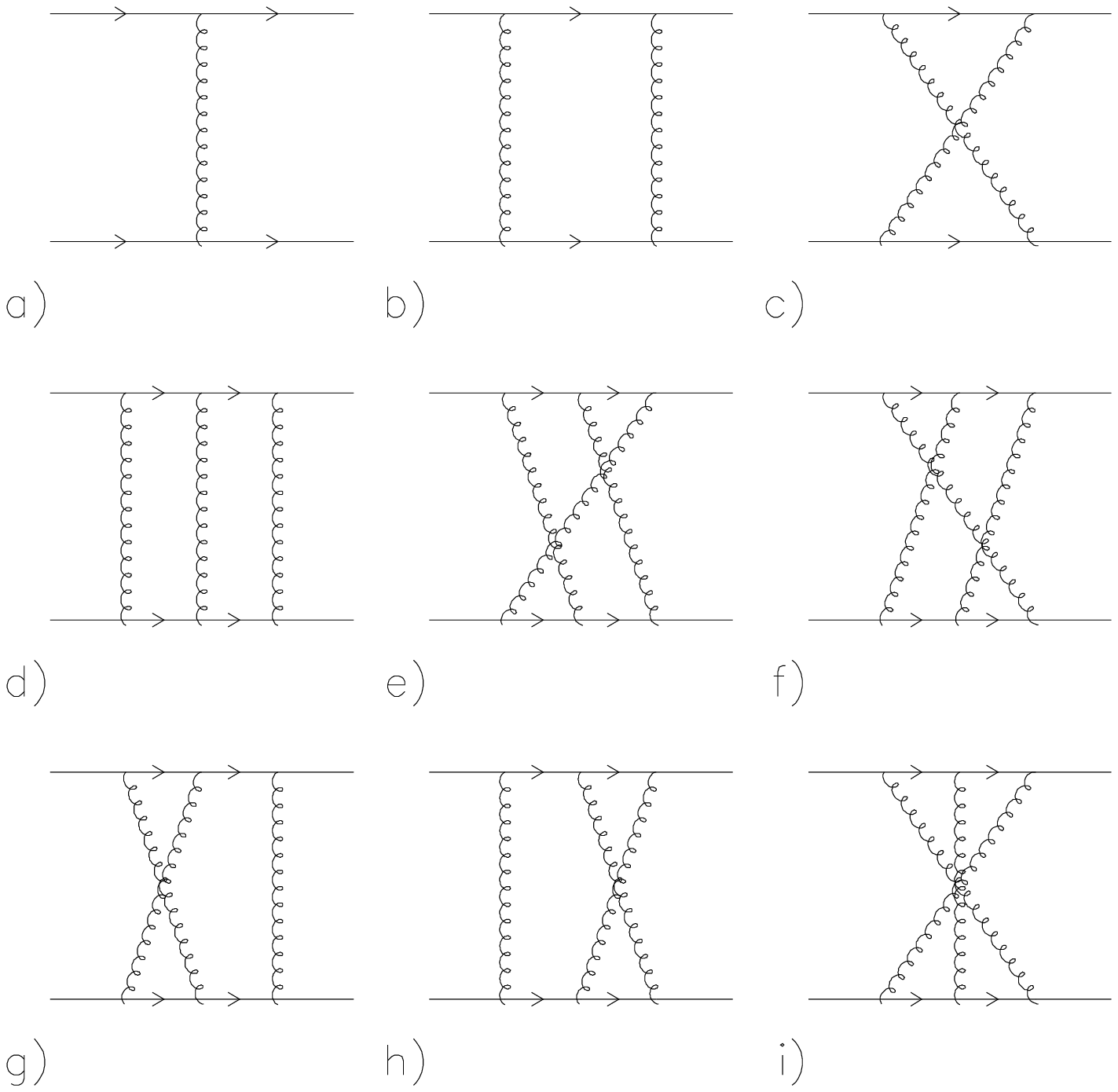}
\label{fig5}
\end{figure}
\vskip 8.5 cm
\begin{itemize}
\item [\bf Fig.~5.] The Feynman diagrams with exchange of one, two and three
gluons which contribute to the quark--quark and quark--antiquark correlators.
\end{itemize}

\newpage

\end{document}